\documentclass[fleqn,usenatbib]{mnras}
\pdfoutput=1

\usepackage{newtxtext}

\usepackage[T1]{fontenc}
\usepackage{ae,aecompl}

\usepackage{graphicx}	
\usepackage{amsmath}	
\usepackage{amssymb}	
\usepackage[british]{babel}             
\usepackage{bm}
\usepackage{physics}
\usepackage[dvipsnames]{xcolor}
\usepackage{orcidlink}
\usepackage{booktabs}
\usepackage{makecell}
\usepackage{ulem}
\usepackage{cleveref}

\usepackage[nolist]{acronym}
\newacro{cpd}[CPD]{circumplanetary disk}
\newacro{RSDs}{redshift space distortions}
\newacro{RSD}{redshift space distortions}
\newacro{DGP}{Dvali-Gabadadze-Porrati}
\newacro{CMB}{cosmic microwave background}
\newacro{SM}{Streaming Model}
\newacro{GSM}{Gaussian Streaming Model}
\newacro{STSM}{Skew-T Streaming Model}

\newacro{LPT}{Lagrangian Perturbation Theory}
\newacro{CLPT}{Convolutional Lagrangian Perturbation Theory}
\newacro{GR}{General Relativity}
\newacro{MG}{Modified Gravity}
\newacro{ST}{skewed student-t}
\newacro{PDF}{Probability Distribution Function}
\newacro{FFT}{Fast Fourier Transform}
\newacro{HOD}{Halo Occupation Distribution}
\newacro{SHAM}{subhalo abundance matching}
\newacro{HMF}{halo mass function}
\newacro{LSS}{large-scale structure}
\newacro{TPCF}{two-point correlation function}
\newacro{EFT}{effective field theory}

\newcommand{\Msun}[0]{{M_\odot}}

\title[Clustering of ELGs and LRGs in modified gravity]{Galaxy clustering in modified gravity from full-physics simulations. I: two-point correlation functions}

\author[Collier, Bose \& Li]{Michael Collier \orcidlink{0000-0001-5628-3837}$^{1}$\thanks{E-mail: michael.collier@durham.ac.uk}, Sownak Bose \orcidlink{0000-0002-0974-5266}$^{1}$ \thanks{E-mail: sownak.bose@durham.ac.uk}, Baojiu Li \orcidlink{0000-0002-1098-9188}$^{1}$\thanks{E-mail: baojiu.li@durham.ac.uk}\\
$^{1}$Institute for Computational Cosmology, Department of Physics, Durham University, South Road, Durham DH1 3LE, UK
}
\date{Accepted XXX. Received YYY; in original form \today}

\pubyear{2024}

\begin{document}
\label{firstpage}
\pagerange{\pageref{firstpage}--\pageref{lastpage}}
\maketitle

\begin{abstract}
We present an in-depth investigation of galaxy clustering based on a new suite of realistic large-box galaxy-formation simulations in $f(R)$ gravity, with a subgrid physics model that has been recalibrated to reproduce various observed stellar and gas properties. We focus on the two-point correlation functions of the luminous red galaxies (LRGs) and emission line galaxies (ELGs), which are primary targets of ongoing and future galaxy surveys such as DESI. One surprising result is that, due to several nontrivial effects of modified gravity on matter clustering and the galaxy-halo connection, the clustering signal does not depend monotonically on the fifth-force strength. For LRGs this complicated behaviour poses a challenge to meaningfully constraining this model. For ELGs, in contrast, this can be straightforwardly explained by the time evolution of the fifth force, which means that weaker $f(R)$ models can display nearly the same---up to $25\%$---deviations from $\Lambda$CDM as the strongest ones, albeit at lower redshifts. This implies that even very weak $f(R)$ models can be strongly constrained, unlike with most other observations. Our results show that galaxy formation acquires a significant environment dependence in $f(R)$ gravity which, if not properly accounted for, may lead to biased constraints on the model. This highlights the essential role of hydrodynamical simulations in future tests of gravity exploring precision galaxy-clustering data from the likes of DESI and \textit{Euclid}.
\end{abstract}




\section{Introduction}
\label{sec:intro}

The current standard model of cosmology, $\Lambda$CDM, asserts the existence of cold dark matter (CDM) and that gravity acts according to Einstein's General Relativity (GR), with a cosmological constant, $\Lambda$, explaining the observed accelerated expansion of the late Universe \citep{1999ApJ...517..565P,1998AJ....116.1009R}. We are now in an era of precision cosmology, with large-scale galaxy surveys such as the Dark Energy Spectroscopic Instrument (DESI) \cite{DESI_EDR_2023} and \textit{Euclid} \citep{Gabarra_EP31} measuring galaxy properties, distribution and clustering, including the baryon acoustic oscillation (BAO), with unprecedented details, allowing us to test any proposed cosmological model on the largest scales. Cosmological parameters have also been tightly constrained through cosmic microwave background (CMB) measurements such as those of \textit{Planck} \citep{Planck15Overview:2016A&A...594A...1P} and WMAP \citep{Bennett_2013}. The values of these parameters are broadly consistent with high-precision measurements of weak gravitational lensing, e.g., Dark Energy Survey \citep[DES,][]{DES_1}, Hyper Suprime Cam \citep[HSC,][]{HSC_1} and Kilo-Degree Survey \citep[KiDS,][]{KiDS_1}, strong lensing, galaxy clusters, and various other observations, despite the emerging ``cosmic tensions" between several measurements \citep[e.g.,][]{2024arXiv240612106E}. Aside from these, GR itself has been tested in new regimes with observations of gravitational waves \citep[e.g.,][]{Abbott_et-al_2016}.

Despite these successes, there are several theoretical problems suggesting $\Lambda$CDM may not be the final answer. One such issue is a lack of a natural physical explanation for $\Lambda$, often understood as the vacuum energy density: its value has frequently been compared to the predicted value of the zero-point energy in quantum field theory, which is many orders of magnitude larger. An alternative to explain this discrepancy is to propose the existence of a dark energy fluid, the equation of state of which has been constrained to be close to $w=-1$. For example, assuming a flat universe and combining data from Supernova Legacy Survey 3 (SNLS3), BAO and WMAP7 has been shown to give the constraint $w = -1.068^{+0.080}_{-0.082}$ \citep{Sullivan_2011}; more recently, using DESI BAO measurements combined with the CMB and type 1a supernova data, this constraint has been further tightened to $w = -0.997 \pm 0.025$ \citep{desicollaboration2024}.

Another interesting possibility is that on the largest scales gravity may behave differently to GR, which causes the Hubble expansion rate to accelerate. Theories of modified gravity (MG) also offer a systematic way to investigate and constrain possible deviations from GR, and as such studies of these MG theories in a cosmological context can serve as a useful test of the latter. While many of the most stringent constraints on MG theories are from the Solar System (and more recently from the use of gravitational waves), tests of these theories in cosmology focus on data from completely different length scales, and are hence complementary. The MG models often lead to different expansion histories and large-scale structure (LSS) to GR, although the models of primary interest here are those with similar expansion history to GR. MG models are typically formed by making some modification to the Einstein-Hilbert action; some common examples of this are DGP braneworld \citep{DvaliDGP:2000PhLB..485..208D}, K-mouflage \citep{Babichev:2009IJMPD..18.2147B}, Gallileon models (e.g., scalar, \citealt{Nicolis_2009}, \citealt{Deffayet_2009} or vector, \citealt{Heisenberg_2014}, \citealt{Beltr_n_Jim_nez_2016}) and $f(R)$ gravity \citep{DeFelice:2010LRR....13....3D,Sotiriou_2010}. Viable MG models often have some kind of screening mechanism to suppress modified gravity effects in certain regimes. In this work we will be interested in $f(R)$ gravity, which exhibits the so-called chameleon screening that can help the model to evade solar system tests of gravity by suppressing MG effects in high density regions \citep[e.g.,][]{Khoury_2004b,Khoury_2004,2006PhRvL..97o1102M,2008PhRvD..78j4021B}.

LSS is often probed with clustering statistics, which tell us the excess probability of finding galaxies in a given spatial configuration. A commonly-used statistic is the two-point correlation function (2PCF), which quantifies the excess probability of finding two galaxies with a given separation, $r$. Higher-order (i.e., $n$-point) clustering statistics are also of interest though generally less well studied and so shall not be part of our focus in this paper. In large cosmological surveys, we measure the spatial positions of galaxies in redshift space, which are subject to so-called redshift-space distortions (RSD), in which the peculiar velocities of galaxies provide an additional contribution to the cosmological redshift, thereby modifying the inferred galaxy distances. The result of this is that position and velocity information are intermixed along the line-of-sight and the clustering in redshift space becomes anisotropic. This makes RSD a particularly powerful probe of gravity as the real-space clustering and velocity field are tested simultaneously \citep[e.g.,][]{2022MNRAS.514..440R}.

Due to the nonlinear nature of the LSS formation and the inherent nonlinearity of many MG theories, the only way to accurately predict the clustering of galaxies and other tracers of the matter field in these theories, especially in the nonlinear regime, is through cosmological simulations. Simulation codes for MG such as \textsc{ecosmog} \citep{Li_2012}, \textsc{mg-gadget} \citep{Puchwein_2013}, \textsc{mg-glam} \citep{Ruan:2021MGGLAMfR,Hern_ndez_Aguayo_2022} and \textsc{mg-arepo} \citep{Arnold_2019} have proven to be successful if computationally expensive -- the properties and clustering of dark matter haloes in MG have been well studied through dark-matter-only (DMO) simulations \citep[e.g.,][]{Arnold_2016,Reverberi_2019,Ruan_2022}. Such simulations offer valuable insights into the cosmic structure formation which is dominated by dark matter, but they have a serious shortcoming, namely they do not directly predict what are observed in practice -- galaxies and baryonic matter. 

The formation and evolution of galaxies are intrinsically connected to the growth of their host dark matter haloes. For $\Lambda$CDM, there have been many efforts to develop a model of the galaxy-halo connection -- a relation between a galaxy's properties and the properties and evolution of its host halo -- which allows one to ``paint" galaxies onto a DMO simulation as a quick way to generate galaxy mock catalogues. There are several approaches for this, including empirical models such as subhalo abundance matching \citep[SHAM, e.g.,][]{Kravtsov_2004,Vale_2004,Tasitsiomi_2004}, and more physically-motivated ones such as semi-analytic models \citep[e.g.,][]{2006RPPh...69.3101B,Conroy_2009,Wechsler_2018,2018MNRAS.474..492M}. One notable example of galaxy-halo connection is the halo occupation distribution \citep[HOD,][]{2002ApJ...575..587B}, which specifies the mean galaxy occupancy in dark matter haloes as a function of halo mass proxies and sometimes extra parameters. The most basic HOD model assumes that the distribution of galaxies in haloes is only dependent on the halo mass. However, using the purely mass-dependent HOD to populate haloes has been shown to result in $\sim 15\%$ inaccuracies in galaxy clustering, which can be remedied with the introduction of an extra parameter \citep[e.g., accounting for local halo environment,][]{Hadzhiyska_2020}. 

For MG models, our understanding of the galaxy-halo connection is fairly rudimentary. In the past studies have usually assumed the same functional form of HOD as in GR, though this has not been thoroughly verified. The common argument is that the HOD, and its parameters, can be determined by matching certain galaxy observables, such as the projected 2PCF and number density, with observations, and any MG model must be able to achieve such a matching (which is generally the case, because even the basic HOD model proves quite flexible). A risk of this empirical approach, however, is that it neglects possible correlations between the HOD model or its parameters and any new physics in the MG model being considered. For example, as mentioned above, due to assembly bias, the HOD can acquire an environment dependence, and environment dependencies are common in MG models with screening mechanisms. It is therefore possible that an incorrect MG model and an incorrect HOD could give galaxy-clustering predictions consistent with observations, leading to the wrong interpretation that this MG model is supported by data. Indeed, as we shall see below, MG can have a nontrivial effect on the HOD itself, and thus we argue that the HOD parameters should \textit{not} be arbitrarily tuned.

This consideration has motivated the present study, where we will look at the galaxy populations and their properties (such as HOD) through a new suite of full-physics hydrodynamical cosmological simulations which can produce realistically various galaxy and cluster observables. Utilisation of this type of simulations for MG models has been somewhat more limited due to the increased computational cost. Recently, however, there have been attempts to simulate smaller volumes as in \cite{Arnold_2019}. The simulations of \cite{Mitchell_2022} are the largest full baryonic physics MG cosmological simulations to date, with a boxsize of $301.75 \, h^{-1}\mathrm{Mpc}$. Recently we have extended this work by including more variants of MG models. We will use the data from these simulations to explore galaxy clustering in one particular MG model, $f(R)$ gravity, here. Studies for other MG models will be presented in future works.

This paper will be organised as follows. In Section~\ref{sect:theory_method} we will describe the Hu-Sawicki $f(R)$ model (Sect.~\ref{sec:f(R)_overview}), which is the specific and popular version of $f(R)$ gravity, the simulations used in this work (Sect.~\ref{sec:simulations}), the galaxy catalogues to be analysed (Sect.~\ref{sec:galaxy_catalog_generation}), and the clustering observables to be studied (Sect.~\ref{sec:correlations}). In Section~\ref{sec:res} we will present the MG predictions of dark matter halo abundances (Sect.~\ref{subsect:HMF}) and the clustering measurements for two classes of galaxies (Sect.~\ref{subsect:CFS}), in both real (Sect.~\ref{subsect:CFS_real_space}) and redshift (Sect.~\ref{subsect:CFS_redshift_space}) spaces, and identify and explain the deviations from GR. To assist the explanation, we have also measured and shown the HODs of the galaxy populations in various $f(R)$ models in Appendix \ref{appendix:HOD}. Finally, we summarise and conclude in Section \ref{sec:conc_and_disc}.

\section{Theory, Simulations and Methodology}
\label{sect:theory_method}

In this section we briefly describe the MG model, $f(R)$ gravity, the hydrodynamics simulations for this model and the resulting galaxy catalogues to be analysed for the rest of this paper. We keep things short and refer the interested readers to relevant reference papers for more details.

\subsection{$f(R)$ gravity}
\label{sec:f(R)_overview}
In $f(R)$ gravity the $\Lambda$ term of $\Lambda$CDM is substituted with a new term, $f(R)$, which is a new nonlinear function, $f$, of the Ricci curvature scalar, $R$. The derivative $f_R \equiv \mathrm{d}f/\mathrm{d}R$ can be interpreted as a new scalar degree of freedom, often called a scalaron, and as such the resulting MG effects are often described as being due to a fifth force originating from this new field. The screening mechanism from this viewpoint can be interpreted as the scalaron's mass, $m_s$, becoming large in high-density regions, so that the associated force becomes very short-ranged, i.e., exponentially suppressed as $\propto\exp(-m_sr)$.

Mathematically, $f(R)$ gravity is characterised by the modified Einstein-Hilbert action \citep[e.g.][]{CarrollLambda:2001LRR.....4....1C}: 
\begin{equation}
S_{f(R)} = \int \sqrt{-g} (R +f(R) +\mathcal{L}_m)\,\dd^4x,
\end{equation}
where $\mathcal{L}_m$ is the matter Lagrangian density and $g$ is the determinant of the metric tensor $g_{\mu\nu}$. 
Minimising this with respect to variations in $g_{\mu\nu}$, we get the modified Einstein field equations for this model:
\begin{equation}
    \label{eqn:EFE}
    G_{\mu\nu} + f_R R_{\mu\nu} - \left[\frac{1}{2} f(R) - \Box f_{R}\right]g_{\mu\nu} - \nabla_\mu \nabla_\nu f_{R} = 8\pi G T^m_{\mu\nu}.
\end{equation}
Here, $\nabla_\mu$ is the covariant derivative, $R_{\mu\nu}$ is the Ricci tensor, $G_{\mu\nu} \equiv R_{\mu\nu} - \frac{1}{2}Rg_{\mu\nu}$ is the Einstein tensor, $T^m_{\mu\nu}$ is the energy-momentum tensor of matter and $G$ is Newton's constant. 

Taking the trace of Eq.~\eqref{eqn:EFE}, we get:
\begin{equation}
    \label{eq:EOM fR}
    \Box f_R = \frac{1}{3}(R - f_R R + 2f(R) + 8\pi G \rho_m),
\end{equation}
where $\rho_m$ is the density of matter and $\Box\equiv\nabla^\mu\nabla_\mu$. This equation can be considered as the equation of motion for the scalaron. 

For the models of interest in this paper, we have $f(R) \ll R$. Furthermore, we work with the quasi-static approximation which assumes that the time derivatives of $f_R$ can be neglected; this has been shown to be valid for the models used in this paper \citep{Bose_2015}. Using these approximations, we find that in the limit where perturbations in $f_R$ about the background value $\bar{f}_R$ are small:
\begin{equation}
    \label{eq:EOM Phi}
    \vec{\nabla}^2 f_R \approx -\frac{1}{3} a^2 \left[ \delta R +8\pi G \delta \rho_m \right],
\end{equation}
and (working in the Newtonian gauge) that:
\begin{equation}
    \vec{\nabla}^2 \Phi \approx \frac{16  \pi G}{3} \delta \rho + \frac{1}{6} \delta R.
\end{equation}
Eqs.~(\ref{eq:EOM fR}, \ref{eq:EOM Phi}) are respectively the equations of motion for the scalar field, $f_R$, and the Newtonian potential, $\Phi$. Note that the counterpart of the latter in GR would be the Poisson equation:
\begin{equation}
    \label{eq:Poisson}
    \vec{\nabla}^2 \Phi_{\rm GR} \approx 4\pi G \delta \rho_m .
\end{equation}

If perturbations in $f_R$ are sufficiently small, then from Eq.~\eqref{eq:EOM Phi} we have $\delta R/6 \approx 4\pi G \delta \rho_m/3$ and so recover the GR case, Eq.~\eqref{eq:Poisson} in $f(R)$ gravity. However, if we have $|\delta R|  \ll 32 \pi G |\delta \rho_m|$ then we find $\vec{\nabla}^2 \Phi \approx (4/3) \vec{\nabla}^2 \Phi_{\mathrm{GR}}$; in other words, the gravitational potential in the GR case is enhanced by a factor of $1/3$.  An ideal $f(R)$ model would exhibit chameleon screening, which involves choosing a function $f(R)$ which results in GR-like gravity on Solar System scales (or comparable regions where the curvature, $R$, is large) whereas on cosmological scales an enhancement up to a factor of $1/3$ may appear. An example of a model featuring this Chameleon screening is Hu-Sawicki (HS) $f(R)$ model \citep{Hu:2007PhRvD..76f4004H}, in which:
\begin{equation}
    f(R) = -m^2 \frac{c_1 (R/m^2)^n}{c_2(R/m^2)^n + 1}
\end{equation}
where $m^2 \equiv H_0^2 \Omega_{\textrm{M}}$ with $\Omega_{\rm M}$ the matter density parameter today, and $n, c_1, c_2$ are free parameters. We take $n=1$.  In the case where $|R| \gg m^2$ for the whole evolution period of interest, the function $f(R)$ is approximately constant, mimicking a cosmological constant, and the choice $\frac{c_1}{2c_2}m^2 \approx \Lambda$ is thus necessary to match the background expansion history to that of $\Lambda\mathrm{CDM}$. This leaves one free parameter to be specified to fix a model, which will be chosen as $f_{R0}$ as we show next. 

With this choice of $f(R)$, the scalaron field is given by
\begin{equation}
    f_R =\frac{\dd f(R)}{\dd R} = - n \frac{c_1}{c_2^2}\left( \frac{m^2}{R} \right)^{n+1},
\end{equation}
and with our choice of parameters we have 
\begin{equation}
    \left|f_{R0}\right|  = \frac{c_1}{c_2^2}\left(\frac{\Omega_{\textrm{M}}}{\Omega_{\textrm{M}}a^{-3}+4\Omega_{\Lambda}}\right)^2,
\end{equation}
which is the background value of $\left|f_{R}\right|$ at $z=0$. Here $\Omega_\Lambda=1-\Omega_{\rm M}$. We will refer to models using the nomenclature `Fx', corresponding to $|f_{R0}| = 10^{-x}$. Models with larger $x$ (or smaller $\left|f_{R0}\right|$) will be referred to as weaker models (since they represent weaker strengths of the  fifth force) and likewise models with smaller $x$ are stronger models.

\subsection{Galaxy Formation Simulations in $f(R)$ Gravity}
\label{sec:simulations}

We have run a new suite of realistic galaxy formation simulations of $f(R)$ gravity, which extends those reported originally in \cite{Mitchell_2022} by having more variations of the modified gravity parameter $f_{R0}$. The parameter values covered in this suite of simulations are $\log\left(|f_{R0}|\right)=-6.0,-5.5,-5.0,-4.5$ and $-4.0$, and according to the above nomenclature we call these models F6.0, F5.5, F5.0, F4.5 and F4.0 respectively. In addition, we have the $\Lambda$CDM counterpart which is equivalent to $f_{R0}=0$. 

These simulations have been run using a version of the N-body and hydrodynamical simulation code \textsc{arepo} \citep{2010MNRAS.401..791S}, modified to include an $f(R)$ gravity solver \citep[see][for more details]{Arnold_2019}, and full baryonic physics with a recalibrated \citep{Mitchell_2022} IllustrisTNG physics model \citep{Pillepich:2017jle}. The retuning of this subgrid model is so that, with the substantially lower mass and force resolutions ($1136^3$ dark matter particles and $1136^3$ initial gas cells in a cubic box of size $301.75h^{-1}\textrm{Mpc}$) than used in IllustrisTNG, the simulations still agree with calibration data of six observables, including the stellar mass function, stellar-to-halo mass relation, star formation rate density, cluster gas mass fraction, galaxy size and black hole mass. We have found that the changes induced by varying $f_{R0}$ on these observables are much smaller than the uncertainties in the observational data themselves, and therefore decided against retuning the subgrid physical parameters for each value of $f_{R0}$ adopted in this work. 

The cosmological parameters of the simulations are taken from the best-fit Planck cosmology \citep{Planck15Parameters:2016A&A...594A..13P}: $(h,\Omega_{\rm M},\Omega_{\rm B},n_{\rm s},\sigma_8) = (0.6774, 0.3089,0.0486,0.9667,0.8159)$, where $\Omega_{\rm B}$ is the present-day density parameter of baryons, $n_{\rm s}$ is the spectral index of the primordial density power spectrum and $\sigma_8$ is the root-mean-squared of the fluctuations of the matter density field today, smoothed on a scale of $8h^{-1}\textrm{Mpc}$. 

Being full-physics runs, these simulations follow the detailed history of radiative gas cooling, star formation, supernova and black hole feedback, etc., to predict the formation of galaxies inside dark matter haloes. These predicted galaxies, subject to further selection based on specific criteria, will be used for the analysis below. More details of these simulations will be given in a forthcoming paper.

\subsection{Galaxy Catalogues}
\label{sec:galaxy_catalog_generation}

In this work we focus on the clustering of two types of galaxy populations -- luminous red galaxies (LRGs) and emission line galaxies (ELGs) -- both of which are primary targets of DESI. LRGs are typically large galaxies which evolve passively and contain a population of old stars, on the other hand ELGs are typically less massive bluer disk galaxies which are still actively forming stars.

\begin{figure*}
    \includegraphics[scale=0.9]{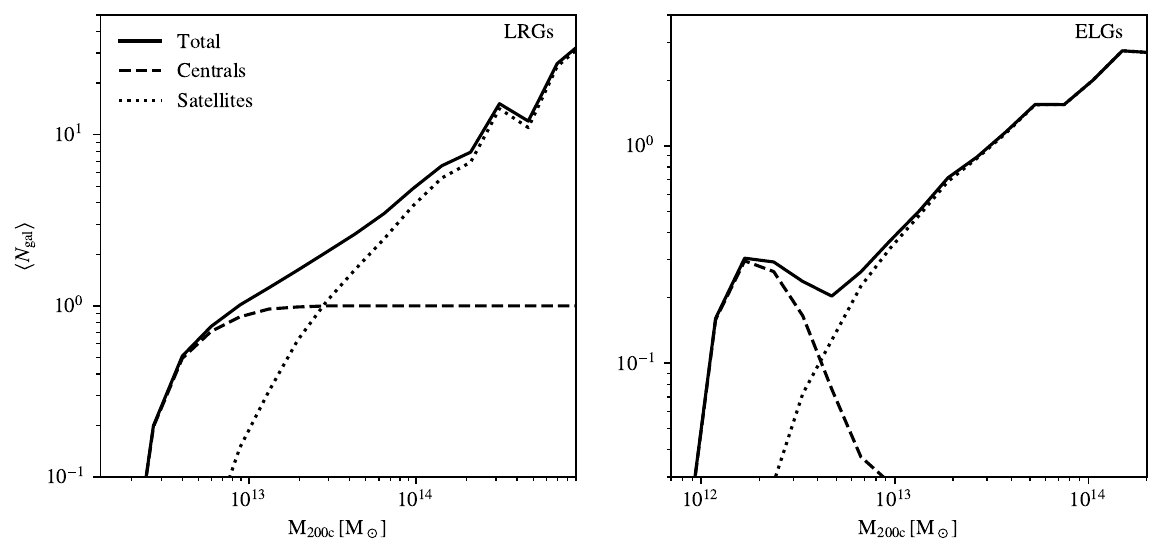}
    \caption{The mean halo occupation as a function of halo mass from our GR LRG (\textit{Left}) and ELG (\textit{Right}) catalogues at number density $n_{\rm g} = 0.001\, h^3\mathrm{Mpc}^{-3}$ and redshift $z = 1.060$ in haloes with evenly spaced bins of logarithmic mass $M_{\rm 200c}$. $M_{\rm 200c}$ is the dark matter mass within $R_{\mathrm{200c}}$ -- the radius from the deepest point in the halo gravitational potential from which the enclosed average density is $200 \times \rho_{\rm c}$, $\rho_{\rm c}$ the critical density of the universe. Central galaxy occupations, satellite galaxy occupations and occupations of all galaxies in the sample (Total) are shown in different line styles as specified in the legend.}
    \label{fig:HOD example}
\end{figure*}

To prepare the LRG catalogues, we rank the simulated galaxies by their stellar mass and take the top $N = n_{\rm g} \times L^3$ galaxies, where $n_{\rm g}$ is the desired target number density of the catalogue. An example of the halo occupancy distribution (HOD) for LRGs is shown in the left panel of Fig.~\ref{fig:HOD example}. Given the strong correlation between the halo and central galaxy stellar masses, and since the LRG sample is selected by stellar mass, we expect that above some mass virtually all haloes should contain an LRG central, i.e., our central occupations for LRGs are essentially 1 above some mass. In Fig.~\ref{fig:HOD example} we see exactly this above around $10^{13} M_{\odot}$, the central occupations rapidly drop to be very close to zero by around $2\times 10^{12} M_{\odot}$, showing we effectively have a mass regime where essentially no haloes have central LRGs. 
Satellite occupations are not restricted to only one per halo as centrals are, and as such their mean occupations can exceed $10$ in larger haloes. However, only larger haloes are typically capable of having a satellite LRG, and due to more numerous small haloes, a majority of our sample will still be central galaxies (e.g., see Fig.~\ref{fig:HMF} later).

In galaxy surveys, ELGs would be selected according to their colour. In simulation data, however, such information is not readily available although it can be estimated with detailed post-processing. A simplified approach has been suggested in \cite{Hadzhiyska_2021}, where it is shown that one can obtain ELG-like galaxy catalogues by ranking the galaxies from the highest specific star formation rate (sSFR) above some mass cut, and and taking the first $N = n_{\rm g} \times L^3$. Due to our relatively low simulation resolution, we have to use a larger stellar mass cutoff of $3 \times 10^{10} M_{\odot}$ for our catalogues than the one used in \cite{Hadzhiyska_2021} in order to have well-resolved objects. Note that such considerations are not needed for LRGs which by their selection process will already have enough star particles.

The right panel of Fig.~\ref{fig:HOD example} shows an example of the HOD for one of our GR ELG samples. An interesting property of the central ELG HOD is that the halo occupation reaches a maximum -- which is less than $1$ -- at a fairly low halo mass, just above $10^{12} M_\odot$ in this case, and monotonically decreases after that. This is because very small haloes generally have not accreted sufficiently dense gas to form stars and seed an ELG, whereas very massive haloes experience significant AGN feedback which heats the gas and prevents further star formation -- they are quenched. In large haloes with quenched central galaxies, it is still possible for ELGs to be seeded as satellites. This can be potentially ascribed to the fact that ELG satellites are on their first passage through their host halo, and have not yet been depleted of their gas reservoir through ram pressure stripping. For ELG samples, unlike with LRG samples, it is not rare to have a halo with a satellite but no central, however a majority of our ELGs will still be central galaxies of small haloes due to these haloes being highly abundant. 

Having a sample of galaxies mainly in small haloes, as in our ELG catalogue, can have certain advantages for testing gravity. This is because in MG models with chameleon screening, in general objects are well screened (i.e., the fifth force suppressed) at early times when the background scalaron value $f_R(a)$ is small, while getting increasingly unscreened at later times. In particular, smaller haloes become unscreened earlier than larger ones, meaning that galaxies in the former would experience the effect of the fifth force longer and therefore present deviations in clustering from GR earlier. This can potentially allow us to see the modified gravity effects in some weak $f(R)$ models which are otherwise hard to distinguish from GR using cosmological observations, and lead to stronger constraints on $f_{R0}$. Another benefit of looking at ELG clustering is that there are more of them than LRGs in cosmological surveys: for example, by its completion, DESI will have measured the spectra of 15.5 million ELGs and 7.5 million LRGs.

\subsection{Galaxy Correlation Functions}
\label{sec:correlations}

Galaxies are biased tracers of the underlying matter distribution, with different types of galaxy forming in different environments. As such, we can use galaxy clustering as a window to some aspects of large-scale structure. One way galaxy clustering can be measured is with correlation functions, which quantify the likelihood of finding a set of galaxies in a given spatial configuration. The simplest version of this is the two-point correlation function, which measures the excess probability of finding two tracers, e.g., galaxies or simulation particles, with a given separation vector. Higher order (e.g. three-point) correlations are also used, though less commonly due to it being more computationally expensive and more challenging to predict theoretically. 

The two-point correlation function is defined as:
\begin{equation}
    \label{eq:natural estimator}
    \xi(\vec{r}) = \langle \delta(\vec{r} +\vec{x})\delta(\vec{x})\rangle_{\vec{x}},
\end{equation}
where $\delta(\vec{x})$ is the density contrast of the tracer under consideration at location $\vec{x}$ and the angular brackets $\langle\cdots\rangle_{\vec{x}}$ denotes the ensemble average, which in a simulation or observation reduces to a simple average over space. 

As we analyse a cubic, periodic simulation box at a single redshift, with no complicated geometry, we will use the following estimator to evaluate the correlation function:
\begin{equation}
    \xi(\vec{r}) = \frac{\mathrm{GG}(\vec{r})}{\mathrm{RR}(\vec{r})} - 1,
\end{equation}
where $\mathrm{GG}$ is the number count of galaxy-galaxy pairs separated by a vector $\vec{r}$ and $\mathrm{RR}$ is random-random pair counts which are computed analytically. This is done by simply considering the number of galaxies within a bins range of a given galaxy to be the proportional the volume of the bin multiplied by the number density (and including a corrective factor of $(N_{\rm g}-1)/N_{\rm g}$ where there are $N_{\rm g}$ galaxies in the sample).

The same is used to calculate the correlation functions in redshift space, which is more relevant given that in real observations the three dimensional distances of galaxies are not directly measurable but are instead inferred from their redshifts. Real galaxy clustering measurements are subject to redshift space distortions (RSD). The two significant RSD effects are the finger-of-god (FoG) effect and the Kaiser effect. In the FoG effect, the orbital motions of galaxies with respect to the host halo center results in a redshift space galaxy distribution which is elongated along the line-of-sight in galaxy clusters. The Kaiser effect is caused by the infall of galaxies into clusters and groups, which makes them appear closer to the cluster center in redshift space.  The relationship between the real ($\vec{r}$) and redshift ($\vec{s}$) space coordinates is:
\begin{equation}
    \vec{s} = \vec{r} + \frac{\vec{v}\cdot \hat{n}}{aH}\hat{n},
\end{equation}
where $\vec{v}$ is the galaxy's peculiar velocity, $H=H(z)$ is the Hubble parameter at redshift $z$ and $\hat{n}$ denotes the unit vector in the line-of-sight direction. The effect of RSD on the correlation function is symmetrical both along the line of sight and perpendicular to the line of sight, and as such we can map all of our points into one quadrant, e.g., $0 \leq \theta \leq \pi/2$, and still fully characterise the behaviour while reducing fractional shot noise.

In real space, the correlation function only depends on the magnitude of the separation vector, $r=|\vec{r}|$, thanks to the statistical isotropy of the large-scale galaxy distribution. In redshift space, however, the galaxy distribution is anisotropic because $\vec{s}$ differs from $\vec{r}$ only in the line-of-sight component, and so we write the correlation function therein as $\xi_s = \xi_s(s,\mu)$ where $s=|\vec{s}|$ and $\mu = \mathrm{cos}\theta$ with $\theta$ the angle between the galaxy-pair separation vector and the line-of-sight vector $\hat{n}$. One way we can characterise these RSD effects is by looking at the multipole decomposition of the correlation function:
\begin{equation}
    \xi_s(s,\mu) = \sum_{i=0}^\infty \xi_l(s) L_l(\mu),
\end{equation}
where $L_l$ is the $l$th order legendre polynomials and $\xi_l(s)$ is the corresponding multipole moment. Due to the symmetries of $\xi_s(s,\mu)$, only even multipoles are nonzero, and so the first two multipoles of interest are the monopole ($l=0$) and the quadrupole ($l=2$). The monopole is essentially the correlation function averaged over $\mu$, which tells us about the number of galaxy pairs expected at a given redshift-space separation, $s$. The quadrupole, roughly speaking, tells us about how much the correlation function is elongated or squashed along the line-of-sight axis: positive values indicate elongations 
(e.g., the FoG effect) and negative values indicate squashing (e.g. the Kaiser effect).

While the two-point correlation function and its multipoles provide useful information that can be used to distinguish between different cosmological models, as we shall find out below, their behaviour is quite complicated with detailed scale dependence. This may pose a significant challenge to practically constraining the models using observational data. As are also interested in comparing the overall behaviour of the correlation function, it will prove useful to reduce its value over a range of scales into a single number, to facilitate more summative comparison of different gravity models at different redshifts. For this purpose, we propose the following volume-averaged correlation function:
\begin{equation}
    \label{eq:scale cf}
    \bar{\xi} = V^{-1}\int^{r_{\mathrm{max}}}_{r_{\mathrm{min}}} \frac{\mathrm{DD}(r)}{\mathrm{RR}(r)} 4 \pi r^2 \, {\dd r} - 1,
\end{equation}
where $V = \int^{r_{\mathrm{max}}}_{r_{\mathrm{min}}} 4 \pi r^2\dd r$ is the volume of the region between radii $r_{\mathrm{min}}$ and $r_{\mathrm{max}}$. We will use this measure for the large scales, with $r_\mathrm{min} =5\,\mathrm{Mpc}$ and $r_\mathrm{max} =20\,\mathrm{Mpc}$. We do not attempt to summarise the small scale behaviour of the correlation function since the correlation function measurement is less reliable and more subject to small scale inaccuracies. 

We estimate the standard error on the correlation function using jackknife resampling. For each sample, the box is divided into $3^3 = 27$ equal cubic sub-volumes, and 27 resamples are generated by removing one of the sub-volumes and computing the correlation function of the galaxies in the remaining volume. The standard error is then obtained as the standard deviation of the correlation function estimate of these resamples.

\section{Results and Discussion}
\label{sec:res}

In this section, we first present the halo mass function (HMF) which will be important in understanding the effect of MG on the dark matter haloes present. We will then show real-space correlation functions, for ELGs and then LRGs, and analyse the differences in clustering between these models. For ELGs we also show the large-scale correlation function, $\bar{\xi}$, as it is useful for showing the model differences at a wider range of redshifts. We will then show the redshift-space correlation functions and discuss the model differences in the ELG clustering and then LRG clustering.

\begin{figure}
    \centering
    \hspace*{-0.5cm}
    \includegraphics[scale = 0.92]{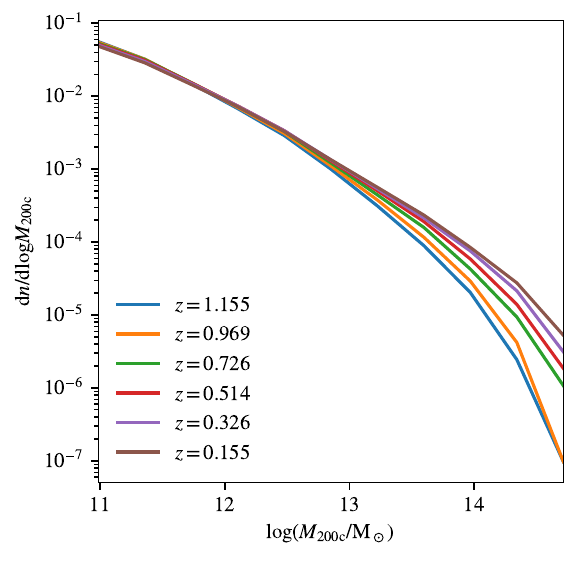}
    \caption{The differential halo mass function for GR at a selection of redshifts. This has been computed by binning haloes in 9 bins of even width of halo mass, $\log(M_{200\rm c}/\Msun)$, from $10^{10.8} \,\Msun - 10^{14.9} \,\Msun$. Different redshift curves are designated with different colours as described in the legend in the lower left of the plot. Note colours do not distinguish different gravity models as will be the case later in the paper as.}
    \label{fig:GR HMF evolution}
\end{figure}

\subsection{Halo Mass Functions (HMF)}
\label{subsect:HMF}

\begin{figure*}
    \centering
    \hspace*{-1.7cm}
    \includegraphics[scale = 0.94]{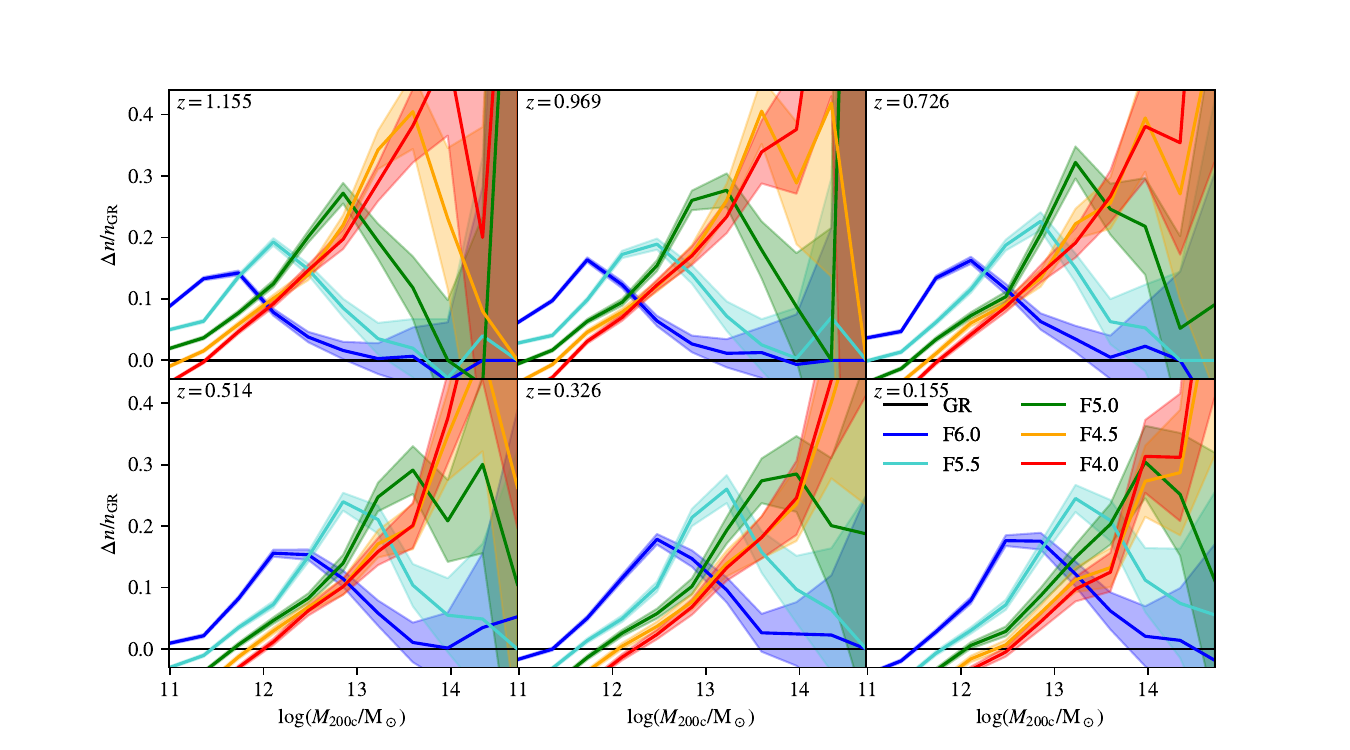}
    \caption{Lines show the fractional difference between the MG HMF and the GR HMF, the shaded regions show the estimated Poisson error on this HMF measurement.The fractional difference in each panel is shown at the redshift specified in the top left corner of each panel. The mass is computed as the dark matter mass within $r_\mathrm{200c}$ - the radius from the center of mass within which the average density is $200 \times \rho_{\rm c}$, the critical density.}
    \label{fig:HMF}
\end{figure*}

Since it will be helpful for us to interpret some of the results to be presented below, we first show the differential HMF, $\dd n/\dd\log M$, for GR at several redshifts in Fig.~\ref{fig:GR HMF evolution}. As expected, the HMF on the high-mass end increases over time, indicating a gradual build-up of these massive objects through accretion and mergers. On the low-mass end, we see fewer objects remain at lower redshifts, showing that these objects have either grown into larger ones, or have been absorbed through mergers. Note that we only show haloes down to a mass of $10^{11} \,\Msun$, as even smaller haloes are unlikely to seed an LRG or ELG and so this regime is not of interest.

We present the fractional difference between the HMFs in our MG models from GR in Fig.~\ref{fig:HMF}, with each panel showing a different redshift as indicated in the legend. Each model is represented by a different colour specified in the legend of the lower right panel. Shaded regions show the $1\sigma$ uncertainty of the HMF measurement which is estimated assuming Poisson noise. These errors are much larger towards the high-mass regime due to the rarity of these haloes within the finite box size of our simulations. We observe that, for most MG models plotted in each panel, there is a ``bump'' in the HMF enhancement which peaks within some mass range. The full-width half maximum of this ``bump'' is around $1.0$--$1.5$ decades in halo mass in most cases, with stronger gravity models having their HMF excesses at higher mass: for example, at $z = 1.155$,  the enhancement in the F60 model peaks at around $5\times 10^{11} \, \Msun$ while for F4.5 it peaks at around $5\times 10^{13} \, \Msun$. At each redshift, stronger gravity models exhibit larger excesses in the HMF with F6.0 peaking at around a $15\%$ excess and F4.5 peaking at around a $40\%$ excess at this same redshift. Furthermore, the height of the HMF excess seems to grow a little across the time period plotted, $1.155\geq z\geq 0.155$. For example, F6.0 only grows from about $15\%$ peak excess to about $20\%$. This growth in fact equates to fewer total haloes in excess of GR: back to the example of F6.0, from Fig.~\ref{fig:HMF} we can see that from $z=1.155$ to $0.155$ the peak of the excess moves from about $10^{11.5} \, \Msun$ to $10^{12.5} \, \Msun$, in which mass range the amplitude of the HMF decreases by a factor of $\sim 10$ (Fig.~\ref{fig:GR HMF evolution}) -- a much more impactful change than the roughly $30\%$ increase in the HMF excess.

\begin{figure*}
    \centering
    \hspace*{-0.3cm}    \includegraphics[scale = 0.76]{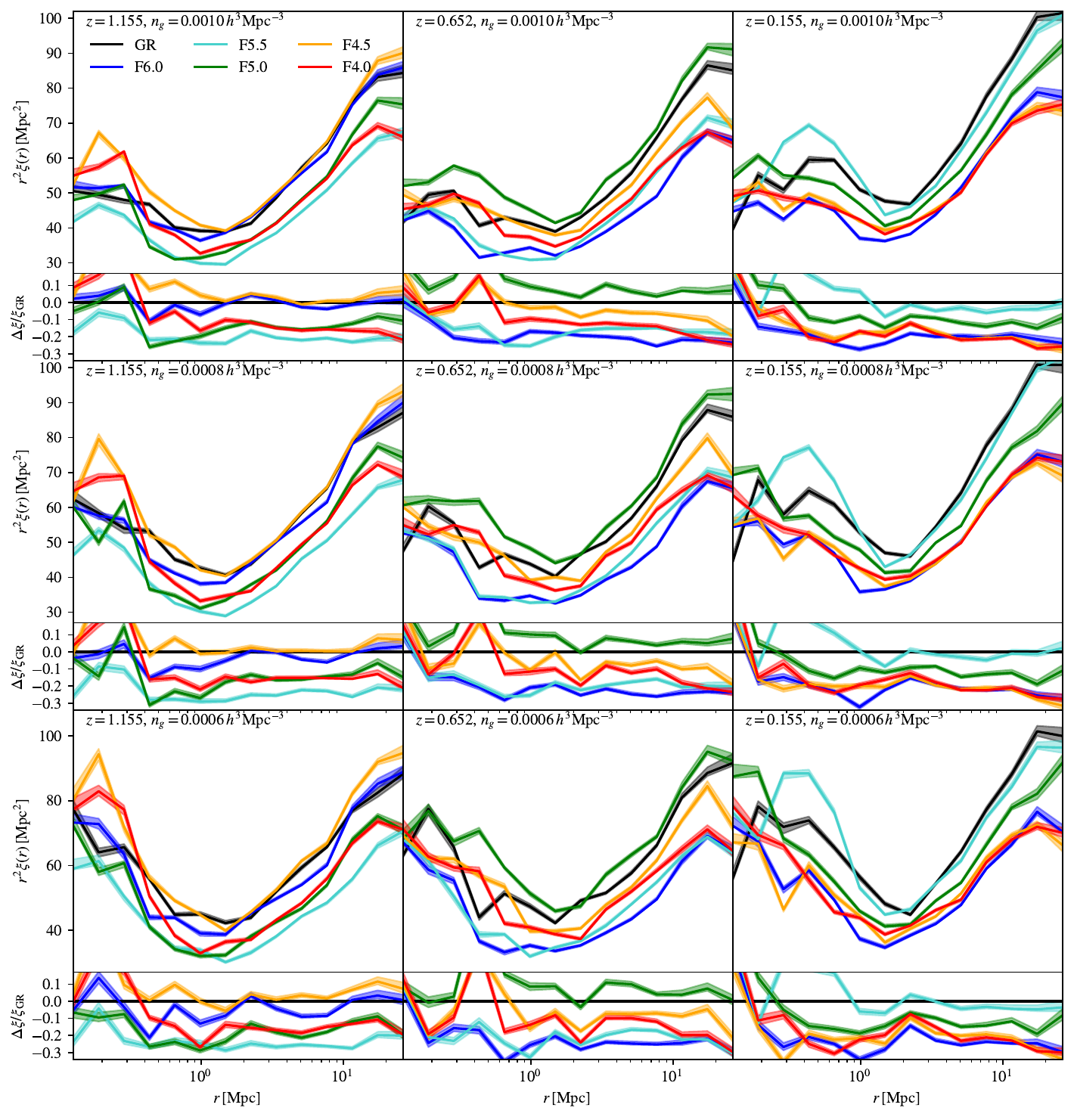}
    \caption{
    \textit{Large Panels:} The real-space correlation function of ELGs multiplied by $r^2$, for different gravity models. \textit{Small Panels:} The fractional difference between the MG and GR real-space correlation functions. Each pair of panels---the large panel and the corresponding small panel immediately below---is for a different (redshift, number density) pair, specified in the upper left corner of the large panel. From left to right panels are decreasing in redshift, number density of the panels increases from bottom to top. Shaded regions show jackknife error estimated as described in Section.~\ref{sec:correlations}.}
    \label{fig:ELG real space cf}
\end{figure*}

We note from Fig.~\ref{fig:HMF} that different models have similarly shaped excesses albeit shifted somewhat in redshift, e.g., the F5.5 excess at $z = 0.969$ is very similar in height, width and mass range to F6.0 at $z = 0.326$. In fact it seems that weaker MG models display the same deviations from GR as stronger ones, except that they do so at lower redshift. In other words, MG models---at least those that are similar to the $f(R)$ gravity model---with varying strengths can be thought of as delayed versions of one another.

Haloes below some mass threshold will be called ``fully unscreened", referring to haloes which have a dynamical mass equal to the lensing mass enhanced by approximately the full maximum $1/3$ factor. Conversely, we refer to haloes with dynamical mass approximately equal to the lensing mass as ``fully screened". For chameleon models, lower-mass haloes become unscreened first: in HS $f(R)$ gravity, it has been shown that the mass beyond which haloes become screened depends on $|f_{R}(z)|/(1+z)$ \citep{Mitchell_2018}. We can understand the behaviour of the HMF as follows: fully unscreened haloes have an enhanced rate of accretion of surrounding matter due to their boosted dynamical mass, and so experience an accelerated growth compared to GR. Meanwhile partially or fully screened haloes see less or no such enhancement respectively. As a result haloes will grow faster than comparable haloes in GR for a while until they become large enough in mass to lose the growth enhancement. And as such we have a ``pile up" of grown haloes around the mass region where screening of haloes kicks in.
An associated effect of this is the depletion of the population of smaller haloes, as these haloes have grown out of the smaller mass range or been accreted early. In Fig.~\ref{fig:HMF} it can be seen especially clearly by late times (e.g. $z= 0.155$) all models have  fewer small haloes than GR below some mass. 

Perhaps no less interestingly, we note there is a saturation effect for stronger models, F4.0 and F4.5 HMFs appear to be practically coinciding with one another from $z = 0.726$ to the present day, while still being distinct from all the weaker models at $z = 0.155$. This is because increasing $|f_{R0}|$ has the effect of unscreening the fifth force from an earlier time, but if this happens before the period when haloes experience most of their growth, then the effect simply would not be reflected in the halo growth. This opens an interesting possibility that late into evolution, strong HS $f(R)$ models become indistinguishable and so it is not always the case that a ``stronger" model necessarily displays stronger deviations from GR in real observables. 

To briefly summarise, in this subsection we have observed, or confirmed, several notable features in the behaviour of the HMFs in HS $f(R)$ gravity, which could have nontrivial implications to their testability. First, a weaker $f(R)$ model is a ``delayed'' version of a stronger one, with qualitatively the same behaviour but displayed at a later time. Second, the maximum enhancement of the HMF with respect to GR happens at different halo masses and different time for different values of $|f_{R0}|$, but the size of the maximum enhancement tends to be insensitive to the latter: this indicates that weaker $f(R)$ models can potentially be as easily distinguishable from GR as stronger ones, if one can have access to earlier-time observations. Finally, there is a saturation effect in the sense that further increasing the value of $|f_{R0}|$ doesn't always lead to ever stronger deviations from GR in terms of physical and observable quantities. 

\subsection{Galaxy Correlation Functions (CFs)}
\label{subsect:CFS}

Next, we consider the clustering of ELGs and LRGs identified from our simulations. To explore the impact of different sample selection criteria we will show several number densities and redshifts. While LRGs live in the largest haloes, ELGs are most often identified as central galaxies of haloes of mass $10^{12} - 10^{13}\Msun$. As a result, and following from the observations of the previous subsection, MG models where the largest halos are yet to become unscreened show differences in ELG population while their LRG populations remain similar to those in GR. On this basis, we preempt that weaker models may show stronger deviations from GR in ELG populations than in LRG populations. 

\subsubsection{Real-space Correlation Functions}
\label{subsect:CFS_real_space}

The real-space correlation functions for ELGs are shown in Fig.~\ref{fig:ELG real space cf}. Each large panel has a corresponding smaller panel below it; the galaxy number density and redshift specific to each correlation function measurement are noted at the top of the large panel. The large panels show the correlation function, $\xi$, multiplied by the square of the pair separation, $r$,  from $0.1\,\rm Mpc < r< 30\, \rm Mpc$. The corresponding smaller panels show the fractional difference of the correlation function for each model from GR. Shaded regions show the $1\sigma$ confidence region estimated via jackknife resampling.

We note immediately that the CFs do not show a systematic deviation from GR in order of fifth-force strength as may be naively expected. GR seems to always be among the most clustered models and F4.0 is always among the least clustered. On the other hand, F4.5 jumps from displaying the strongest clustering at $z = 1.155$ to having among the weakest clustering by $z=0.155$. With a few exceptions, ELGs in the MG models are less clustered than those in GR. In F5.5 at $z = 1.155$, F6.0 and F5.5 at $z=0.652$ and F4.0, F4.5 and F6.0 at $z= 0.155$ we note particularly large deviations of $20-30 \%$ below GR at $r \geq 5 \, \rm Mpc$. These large deviations are present also in the weaker gravity models, F5.5 and F6.0. Interestingly, we will see later this is no longer true when we consider the LRG populations in these models. These observations indicate that for ELGs, there is something about the selection of galaxies that is more important than the underlying clustering of haloes, and that these selection factors affect even the weaker MG models. We will show shortly how the large deviations seen in Fig.~\ref{fig:ELG real space cf} propagate into observable measures of galaxy clustering, making ELG clustering an interesting avenue to explore for constraining even weak Chameleon models.

Fig.~\ref{fig:ELG real space cf} shows at fixed redshift, the relative ordering of the models is independent of the number density. For example, the $z = 1.155$ panels all have a fairly consistent ordering on scales $>1\,\mathrm{Mpc}$ (from bottom to top) of F5.5, F4.0, F5.0, GR, F6.0, F4.5. While the range in number densities is not very large, it is apparent that the differences in clustering between models are not particularly sensitive to the choice of number density for ELGs.

On the other hand, the clustering signal depends much more strongly on redshift. We see that the F6.0 correlation function rapidly drops in magnitude from being comparable to GR at $z = 1.155$ to being $25\%$ smaller than GR by $z = 0.652$. The redshift evolution of the correlation function, given a value of $\left|f_{R0}\right|$, can be partly understood based on our earlier discussions regarding screening and the evolution of the HMF. We previously noted that most ELGs are central galaxies of haloes of mass $10^{12}-10^{13} \, \Msun$. When the HMF excess described in Fig.~\ref{fig:HMF} coincides with this range, we expect that these grown haloes will seed ELGs earlier than ungrown counterpart haloes in GR, since the increased gravity should accelerate the accretion of gas and, subsequently, enhance the specific star formation rate early. We see that at around $z = 0.726$--$0.514$ is when the excess in the HMF overlaps this mass region. This coincides with the maximum reduction in the F6.0 case of $25\%$ in the ELG CF seen at $z = 0.652$, c.f.~Fig.~\ref{fig:ELG real space cf}. The clustering is weaker, because the ELG host haloes in F6.0 grow from initially lower density peaks, which are inherently less clustered.

\begin{figure*}
    \centering
    \hspace*{-0.3cm}
    \includegraphics[scale = 1.02]{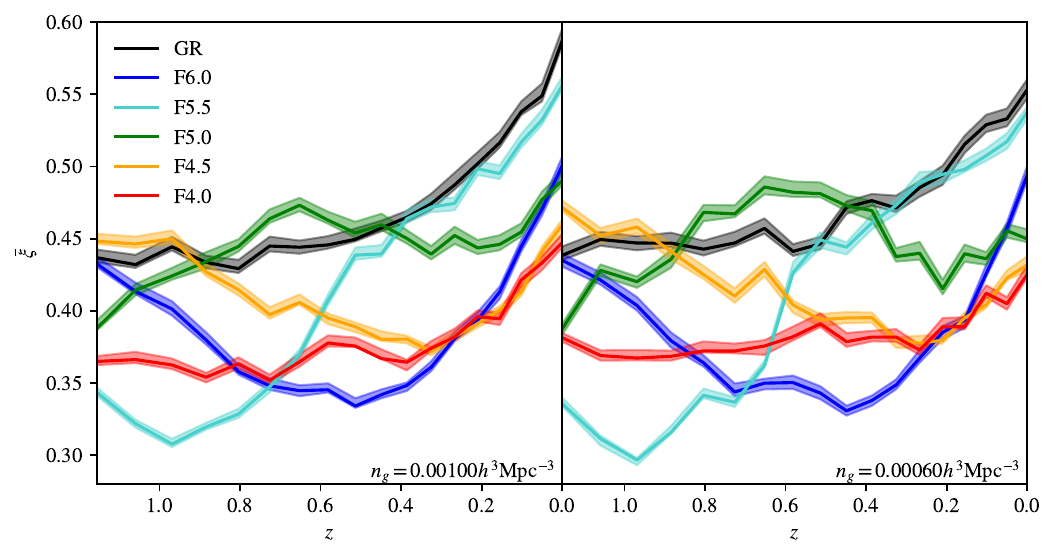}
    \caption{The scale integrated correlation function as described in Eq.~\ref{eq:scale cf} measured for ELG galaxy samples in several different gravity models at two different number densities --  $n_{\rm g} = 0.0010 \, h^3 \mathrm{Mpc}^{-3}$ and $n_{\rm g} = 0.0006 \, h^3 \mathrm{Mpc}^{-3}$ -- plotted against redshift, $z$. For the integration, we have assumed $r_\mathrm{min}= 5\, \rm Mpc$, $r_\mathrm{max} = 20\, \rm Mpc$ computed numerically with $10$ bins. This statistic characterises the correlation function on large scales.}
    \label{fig:ELG LS cf}
\end{figure*}

One difficulty in analysing the evolution of the CF in these MG models from Fig.~\ref{fig:ELG real space cf} is that it only shows 3 different redshifts. We now investigate a wider range of redshifts in order to assess in full detail the evolution of the correlation function. The integrated large-scale correlation function proposed in Eq.~\eqref{eq:scale cf} is shown in Fig.~\ref{fig:ELG LS cf}. Shaded regions show the region of $1\sigma$ confidence, estimated via jackknife resampling. We choose only to show $n_{\rm g} = 0.001 \, h^3 \mathrm{Mpc}^{-3}$, due to our previous observation that number density does not affect the relative model differences. Each model curve in Fig.~\ref{fig:ELG LS cf} is visually distinct. However, note similarities between neighbouring values of $\left|f_{R0}\right|$. Each model at high redshift deviates from GR in the same way as the next most screened model at low redshift, so it appears as if each model's behaviour is not unique but is showing a different part of some more complicated shared pattern of evolution across $1.2<z<0$. For example, the F6.0 curve starts out in agreement with GR before dipping below it---reaching a maximum difference at $z = 0.6$---then gradually returning towards GR again after this. On the other hand, F5.5 starts out below GR reaching maximum difference at $z = 1.0$ and returning to coincide with the GR curve by $z = 0.4$. The common evolution of the ELG clustering can be explained by what we observed above in Fig.~\ref{fig:HMF}, namely weaker MG models can be considered as delayed versions of stronger ones. We also observe the saturation effect here, in agreement with what was previously noted in Fig.~\ref{fig:HMF}: the integrated CFs for F4.5 and F4.0 practically coincide for $z<0.6$, and we can furthermore see the F5.0 integrated CF approaching F4.0 and F4.5 close to $z=0$.

To provide further support to the observations and explanations above, in Appendix \ref{appendix:HOD} we have shown and discussed the behaviour of the HODs for ELGs (Fig.~\ref{fig:ELG HOD}).

\begin{figure*}
    \centering
    \hspace*{-0.3cm}
    \includegraphics[scale = 0.76]{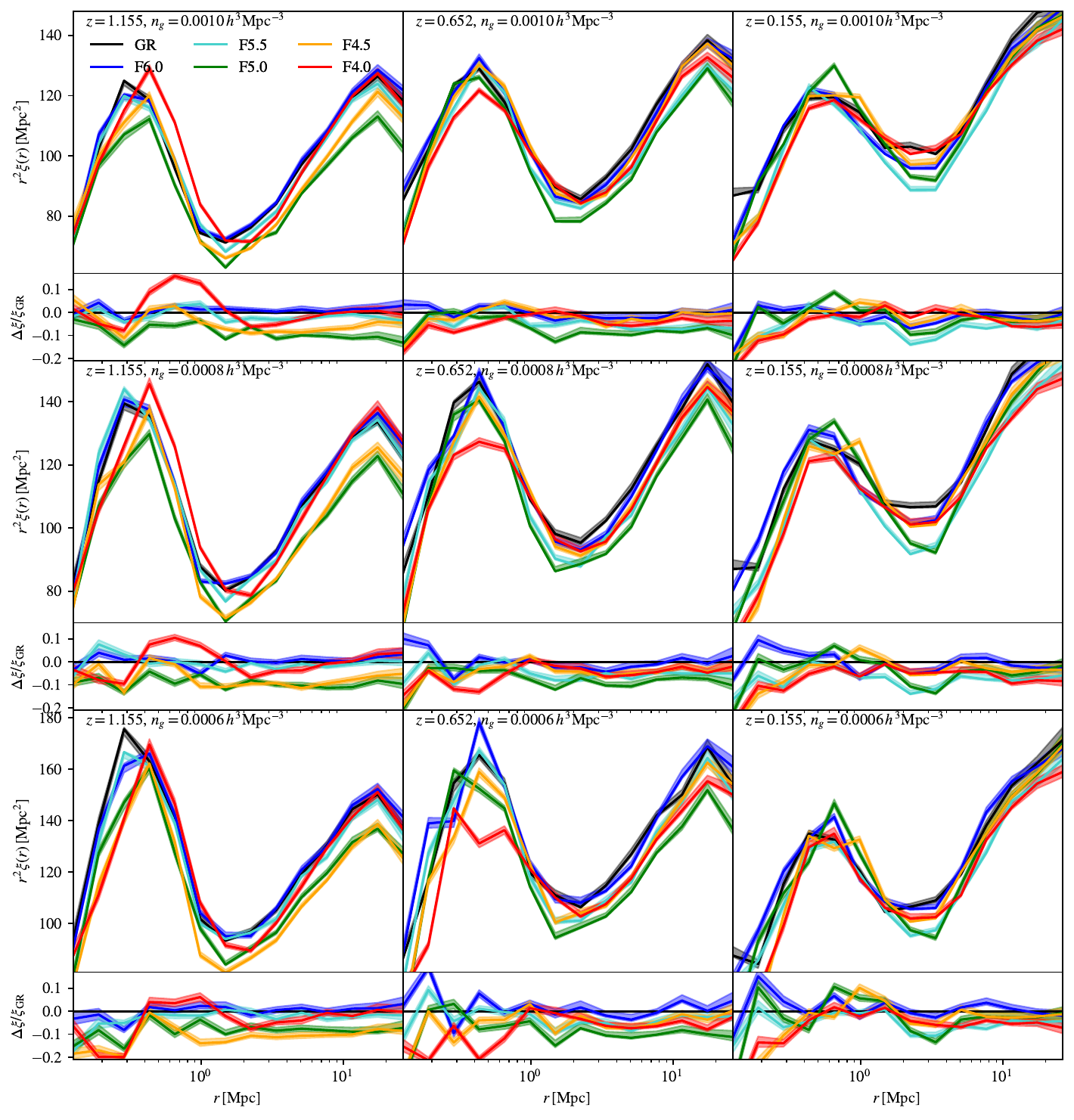}
    \caption{\textit{Large Panels:} The real space correlation function of LRGs multiplied by square separation, $r^2$, for different gravity models. \textit{Small Panels:} The fractional difference between the MG and GR real space correlation function.
    Each pair of panels, large panel and corresponding small panel immediately below, is for a different (redshift, number density) pair, specified at the top of the large panel. From left to right panels are decreasing in redshift, number density of the panels increases from bottom to top. Note that the limits of the vertical axes of different rows of panel pairs differs. Shaded regions show jackknife error estimated as described in Sect.~\ref{sec:correlations}.}
    \label{fig:LRG real space cf}
\end{figure*}

The real-space correlation functions for LRGs are shown in Fig.~\ref{fig:LRG real space cf}. An important observation is that the clustering strength for LRGs is in general larger than for ELGs on the scales shown. This is consistent with the fact that LRG populations exist in the largest haloes, which tend to be more clustered than smaller haloes. The deviations in the LRG CFs for the MG models from GR are at most on the 10\% level; deviations of this magnitude are seen for F5.0 and F4.5 at $z = 1.155$ and F5.0 at $z = 0.652$. This is less than half the largest ELG deviations shown in Fig.~\ref{fig:ELG real space cf}. An interesting observation is that, for all redshifts and LRG number densities shown in Fig.~\ref{fig:LRG real space cf}, both F4.0 (the strongest MG model) and F6.0 (the weakest) show nearly identical clustering strength as GR, while the other, intermediate, models display a stronger deviation.

To better understand this behaviour, we again show the HODs for LRGs in Fig.~\ref{fig:LRG HOD} of Appendix \ref{appendix:HOD}. There, we can observe that the deviation of the HOD from the GR prediction does indeed follow the same order as the fifth-force strength. Therefore, we have two competing effects here. On the one hand, a stronger MG model naturally has stronger and less screened fifth forces, so that for a fixed LRG number density more of these LRGs can be hosted by haloes forming from smaller initial density peaks in low-density environments (which thanks to the unscreened fifth force grow more quickly and become more massive even than some larger initial density peaks in more dense environments). Since lower initial density peaks tend to be less clustered, this leads to a reduction in the LRG clustering. On the other hand, if the fifth force is strong enough, such as in F4.0, it can substantially increase the clustering of haloes and hence the LRGs they host.

According to this picture, F6.0 has nearly identical LRG clustering to GR simply because the fifth force in this model is effectively screened. But for F4.0, the above-mentioned two competing effects balance out, leading also to a clustering signal that is similar as in GR. For the intermediate models, such as F4.5 and F5.0 at $z=1.155$, it appears that the first effect dominates, resulting in an overall weaker clustering than in GR. Again, we note the ``saturation" effect as reflected by the fact that F4.0 and F4.5 have nearly identical clustering signals at $z=0.652$ and $0.155$.

\begin{figure*}
    \centering
    \hspace*{-0.3cm}
    \includegraphics[scale = 0.75]{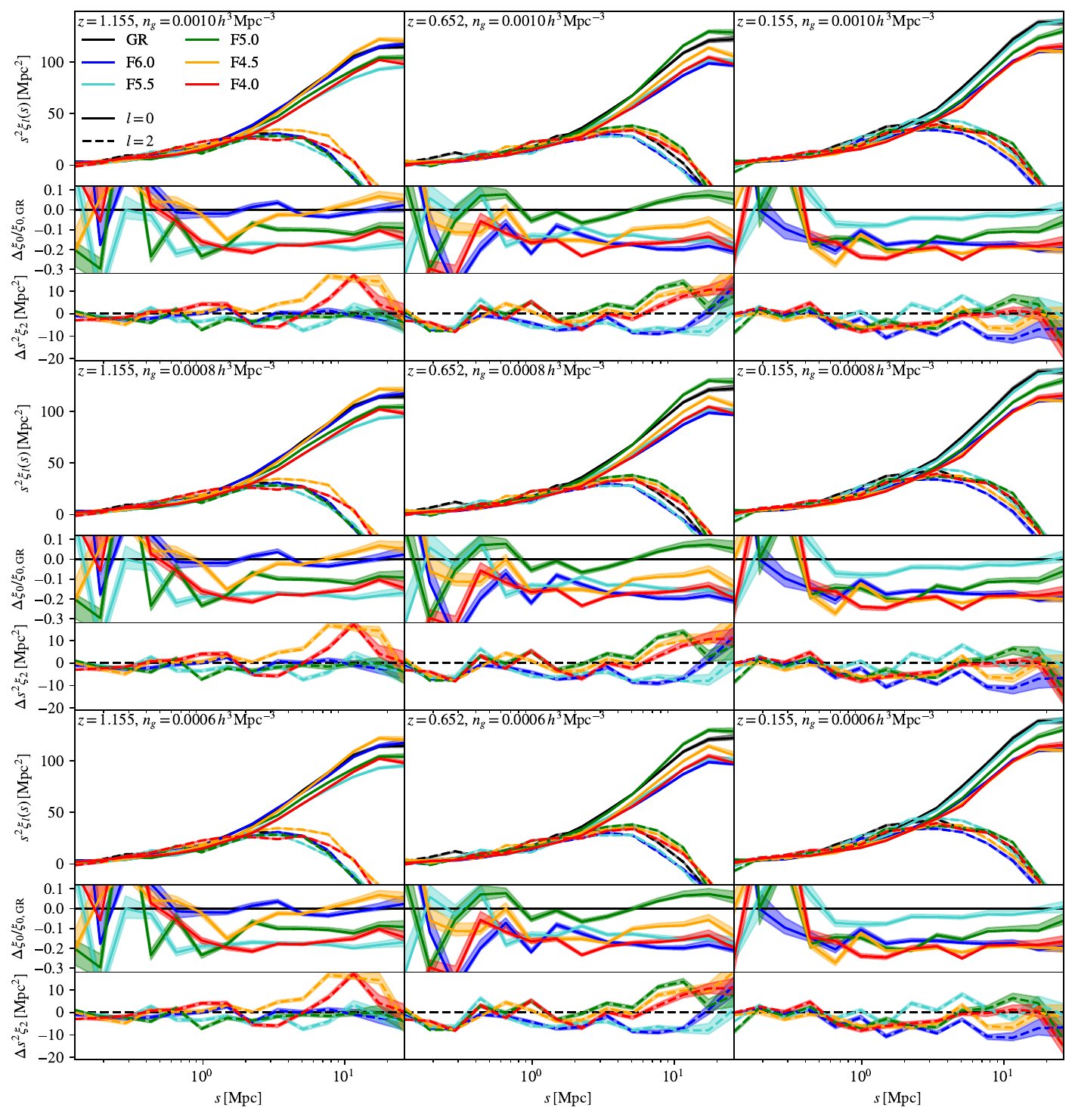}
    \caption{\textit{Large panels:} $l=0$ and $l=2$ multipoles of the redshift space correlation function, distinguished by linestyles as described by the legend. \textit{Upper small panels:} Fractional difference of the monopole in MG from GR. \textit{Lower small panels:} Difference of the quadrupole in MG from GR. All panels are for ELG populations, the large upper panels and corresponding pair of small lower panels are plotted for the redshift and number density specified at the top the upper panel. Shaded regions show jackknife error estimated as described in Sect.~\ref{sec:correlations}. Different models are represented by different colours as described in the legend.}
    \label{fig:multipoles ELG}
\end{figure*}

\begin{figure*}
    \centering
    \hspace*{-0.3cm}
    \includegraphics[scale = 0.75]{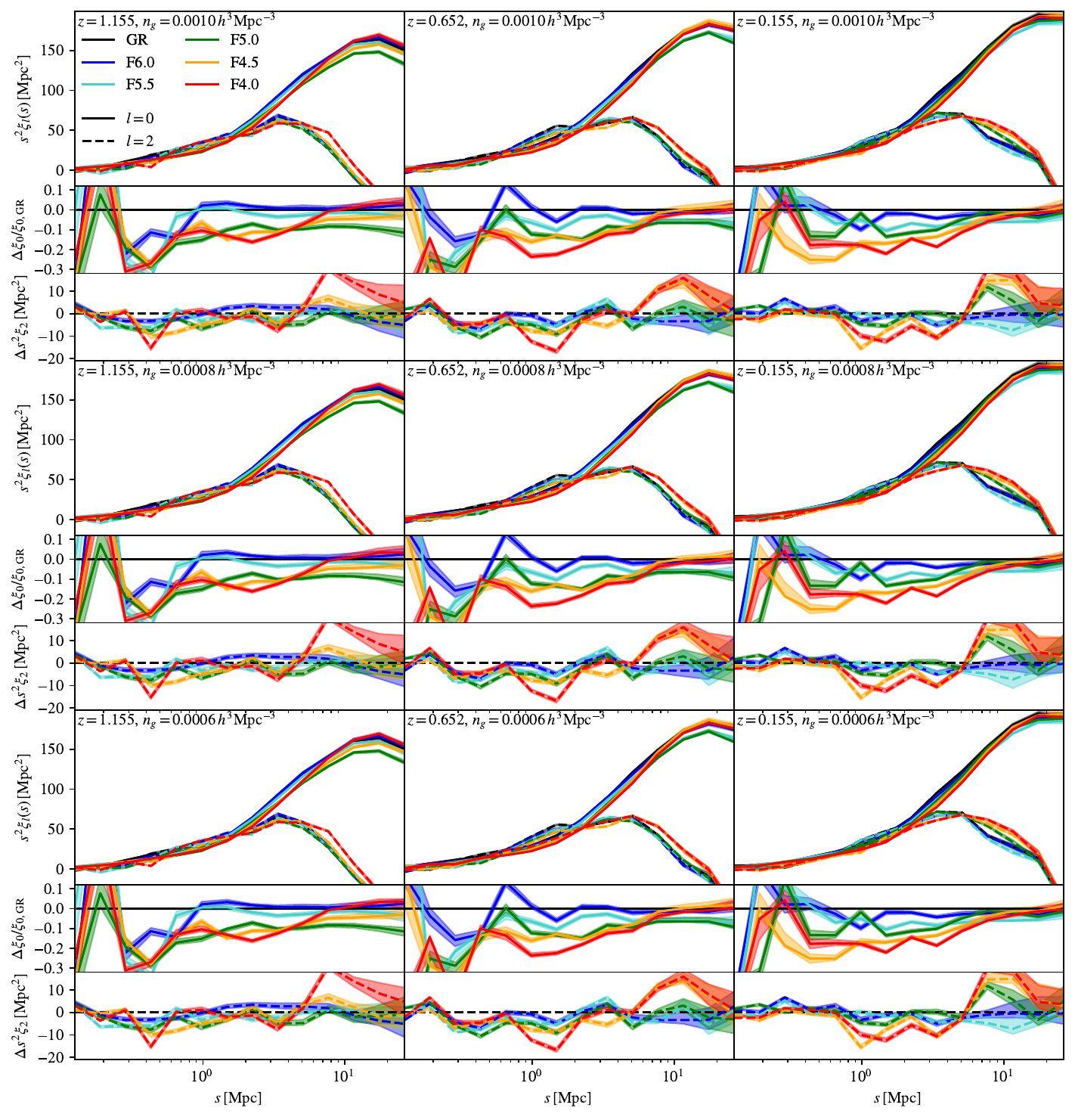}
    \caption{\textit{Large panels:} $l=0$ and $l=2$ multipoles of the redshift space correlation function.
    \textit{Upper small panels:} Fractional difference of the monopole in MG from GR. \textit{Lower small panels:} Difference of the quadrupole in MG from GR. Shaded regions show the estimated jackknife error in the difference. All panels are for LRG populations, the large upper panels and corresponding pair of small lower panels are plotted for the redshift and number density specified at the top the upper panel. Shaded regions show the jackknife error estimated as described in Sect.~\ref{sec:correlations}.}
    \label{fig:multipoles LRG}
\end{figure*}

\subsubsection{Redshift-space Correlation Functions}
\label{subsect:CFS_redshift_space}

We now look at the redshift space CFs for ELGs in Fig.~\ref{fig:multipoles ELG}. This features triplets of panels with a large panel at the top and two corresponding small panels below; each triplet is for the number density and redshift specified at the top of the large panel. The large panels show the monopole $\xi_0$ and quadrupole $\xi_2$ of the correlation function in redshift space, multiplied by the square of the pair separation, $s$ (e.g., $s^2\xi_0(s)$). The panels immediately below the large panels show the \textit{fractional} difference of the monopole of each model from GR, the panels below these show the \textit{absolute} difference of the quadrupole of each model from GR. In each panel, the $1\sigma$ region, estimated with jackknife sampling, is shown by the shaded regions.

We first discuss the monopole. At small scales, e.g., $<1\, \mathrm{Mpc}$, the galaxy pair counts are suppressed by the FoG effect. 
However, in this work we are primarily interested in the large-scale CF. We find a significant fractional difference of $\sim 25\%$ in the MG models deviating most from GR across the entire scale range $>3\,\mathrm{Mpc}$. For example, at $z = 1.155$, F5.5 shows deviations of this size, at $z=0.652$ F5.5 and F6.0, and at $z = 0.155$ F6.0, F4.5 and F4.0. We note that the relative difference between models shows extremely similar behaviour to Fig.~\ref{fig:ELG LS cf} -- in each panel, the models are virtually always in the same vertical ordering as in Fig.~\ref{fig:ELG real space cf}, which is to be expected since the monopole has similar information content the the real space CF, particularly on large scales. As such, our description and suggested explanations of the model differences in Fig.~\ref{fig:ELG real space cf} should be valid for the monopole as well.

The $25\%$ differences from GR seen in the monopole are large, especially when compared to the percent-level expected precision of modern surveys such as DESI. While this is promising, it is important to keep in mind that the galaxy catalogues here are not mock datasets for any real survey. It will be useful for realistic ELG mocks to be produced for ongoing or upcoming cosmological surveys in the future. We could also benefit from having higher-resolution simulations with a higher number of snapshots towards higher redshifts. Presently, due to the previously-described resolution issues with our simulation, ELG samples at e.g., $n_{\rm g} = 0.001\, h^3\mathrm{Mpc}^{-3}$ can only be created below $z = 1.26$; since, for example, DESI will measure the spectra of ELGs for $0.6<z<1.6$, we are unable to study ELGs in the entire volume contained in $1.26<z<1.6$. We plan to revisit this in a future work.

The behaviour of the quadrupole is more complicated, and the ordering with respect to the fifth-force strength is quite different from that of the monopole. Naively, at large enough scales the quadrupole is sensitive to the galaxy pairwise velocity, and hence the velocity field. In particular, it may be expected that the velocity bias is small so that tracer galaxies of different type and haloes of different mass have the same velocity field, which in $f(R)$ gravity can be significantly enhanced \citep[e.g.,][]{2013MNRAS.428..743L,2012MNRAS.425.2128J}. However, it is important to note that the amplitude of the quadrupole also depends on the real-space CF \citep[e.g.,][]{Cuesta-Lazaro:2020ihk}. Such a mixture and competition of effects makes it difficult to interpret the behaviour of the quadrupole, or use it to test/constrain models -- this seems to go against the established wisdom in the field. Another potential complication originates from the fact that in Fig.~\ref{fig:ELG LS cf} we have focused on the CF multipoles on relatively small scales, which may be significantly contaminated by the FoG effect; future larger-box simulations should provide more definitive answers to this question.

We now analyse the redshift-space CFs for LRGs as shown in Fig.~\ref{fig:multipoles LRG}. The monopole amplitude is significantly larger than for our ELG populations, which is another example of the monopole having similar information content to the real-space CF on the large scales. In the scale range of $s>3\, \rm Mpc$, similarly to ELG populations, the model ordering in the monopole of each panel is consistent with the corresponding panel in Fig.~\ref{fig:LRG real space cf}. However, in some cases towards the lower-$s$ end of this regime, the stronger models such as F4.0 have substantially reduced amplitudes of the monopole even though the real-space CF is nearly identical to GR. This can be easily explained, as we expect LRGs to present a greater FoG effect due to haloes hosting multiple galaxies more frequently and   typically also having larger mass. This should move more galaxies to larger separations $s$ in redshift space and so reduce the monopole at low $s$. In contrast to the smaller scales, on large scales we see at most $\sim 10\%$ model differences, consistent with the result for the real-space CF.

Finally, we look at the model differences in the LRG quadrupole. The differences between the last few bins of the quadrupoles in all panels are dominated by the jackknife error and as such we will focus on intermediate scales, $3<s<20\, \mathrm{Mpc}$. Universally, we find that the F4.0 model rises well above GR (typically by $s^2 \Delta \xi_2 \approx 10 \, \mathrm{Mpc}^2$), with other strong models up to F5.0 following suit by $z = 0.155$. This is likely associated with the enhanced velocity field in $f(R)$ gravity. Towards the lower end of this $s$ range, we see a reduction of the quadrupole in the MG models compared to GR, again indicating contamination by the FoG effect.
Meanwhile, the weaker models remain closer to GR, as the large haloes in them remain screened even at low redshift. Despite weaker monopole signature than ELGs on large scales, LRG clustering could still be a useful test of gravity for stronger gravity models. Weak models like F6.0 and F5.5 show generally very small deviations from GR in LRG clustering, both in real and in redshift space. It is interesting that LRGs and ELGs offer very different, and potentially complementary, constraints of chameleon $f(R)$ gravity, demonstrating the value of considering multi-tracer clustering statistics. 

\section{Discussion and Conclusions}
\label{sec:conc_and_disc}

With the launch of various Stage IV cosmological galaxy surveys such as DESI and \textit{Euclid}, galaxy clustering is set to become a major source for cosmological constraints, in particular in the context of MG models. However, questions remain around how reliable these constraints will be, given the uncertainties in our understanding of galaxy formation and the galaxy-halo connection. In our expectation, these ingredients can be strongly affected by any new physics presented in MG models, making it a risky strategy to consider the galaxy-halo connection a free function to be (jointly) determined by observations. 

To address this question, in this work we have run an extended suite of full-physics galaxy-formation simulations in the popular HS $f(R)$ gravity model, and use these to study the HOD and the clustering of two types of galaxies, ELGs and LRGs. The investigation leads to a number of interesting, and to some extent surprising, results, which we summarise below.

First, unlike in some other physical quantities such as the HMF, the deviation in the clustering signals from GR do not always follow the order of the fifth-force strength. This applies to both LRGs and ELGs.

For LRGs, this is because of two competing effects, one being the fact that in MG models more of the LRGs are hosted by haloes that collapse from lower initial density peaks in low-density environments, which are less clustered. The other effect is the overall enhanced clustering of haloes due to the fifth force. The result of this competition is that both F6.0 (the weakest MG model) and F4.0 (the strongest) show nearly identical real-space clustering strength as in GR. Such non-monotonic behaviour inevitably complicates the use of LRG clustering as a testbed for chameleon $f(R)$ gravity (as has been done extensively in the literature so far). 

For ELGs, the result is more interesting: the strongest deviation from GR happens for different MG models at different times. For example, at $z=0.155$ it is the weakest MG model, F6.0, that differs most from GR in the real-space clustering strength, while at $z=0.652$ the model that deviates the most from GR is F5.5. Moreover, the maximum deviation from GR is not sensitive to the value of $\left|f_{R0}\right|$. By looking at the HODs (Fig.~\ref{fig:ELG HOD}), we have provided a physical explanation for this behavior based on how the fifth force affects halo abundances and star formation. While the behaviour here is again non-monotonic with respect to the fifth-force strength, the physics behind it is simpler than in the LRG case. This suggests that by looking at the ``correct" redshift one can potentially use ELG clustering to place stringent constraints on even the weakest $f(R)$ models. These can be considered as the ``sweet spots'' for gravity tests: as mentioned frequently in the preceding discussion, weaker $f(R)$ models can be considered as a delayed version of stronger ones, and each model has its own ``sweet spot" at which we can maximise the potential of constraining it. We also note that the $\simeq25\%$ deviations from GR in even the F6.0 model are quite substantial considering the precision of data expected from the current and upcoming generation of observations.

Second, we have investigated the clustering of LRGs and ELGs in redshift space. For ELGs, the monopole of the CF shows the same behaviour as seen in the real-space CF, confirming that the effect described above can indeed be measured and used in real observations. The quadrupole, however, displays less potential of distinguishing between different MG models, which is a bit surprising because one would naively expect the fifth force to substantially increase galaxy peculiar velocities. This is because the quadrupole depends on both the galaxy pairwise velocity and the real-space clustering, and these two effects cancel out to some extent.

The redshift-space clustering of LRGs again broadly follows the behaviour of the corresponding real-space clustering. However, the FoG effect is much stronger here thanks to the typically higher host-halo masses, and this decreases the monopole between $\simeq1$--$10\,\textrm{Mpc}$. For example, although F4.0 has nearly identical real-space CF as GR and F6.0 in this scale range, it has a significantly lower redshift-space CF monopole, which is another surprise that comes as a convenience (we want to use LRG clustering to constrain models after all!).

Finally, in various statistics examined in this paper we have observed a ``saturation" effect; where the stronger MG models, such as F4.0 and F4.5, have nearly identical behaviour and predictions at late times. This can be roughly understood as follows; the strongest model, such as F4.0, has already maximised its ability of causing deviations from GR well before the present day, and this means that the weaker models, such as F4.5, can have time to ``catch up" at low $z$. We do notice, however, that the saturation happens only at late times, so that higher redshift data can still be used to distinguish between these models.

One of the commonly-used galaxy clustering observables that we have not included in this study is the projected 2PCF. This is because our limited box size means that we cannot measure the real- or redshift-space CFs to galaxy separations above $\simeq20$--$30\,\textrm{Mpc}$, and therefore we cannot perform the line-of-sight projection reliably. But given the results above we would expect similar levels of deviations from GR in both. This further highlights our argument above that it is risky to fix the HOD parameters empirically by matching the predicted projected 2PCF with the observed one, and that a detailed study of how MG affects galaxy formation is essential if we wish to test the models using galaxy clustering.

To study the projected 2PCFs, we will need larger simulations, which will be a plan for future. Items we will consider in forthcoming work include (i) running simulations at higher resolution to allow lower stellar mass cuts when selecting the ELGs; (ii) a more realistic selection of ELGs based on (observable) galaxy colours and magnitudes \citep[e.g.,][]{2022MNRAS.512.5793Y}; and (iii) making realistic mock galaxy catalogues for future observations, by taking into account effects such as lightcone, survey geometry and redshift distribution. We will also study other classes of MG models in future works.

To summarise, galaxy clustering is a very promising avenue for testing gravity models in cosmology, but perhaps in different ways from what was believed previously. With the rise of large realistic hydrodynamical simulations, we have entered a new era where detailed galaxy properties, including their spatial distribution, can be predicted by such simulations. This will offer tantalising new opportunities for cosmological tests of gravity.

\section*{Acknowledgements}

MC is supported by a UK Science \& Technology Facilities Council (STFC) PhD studentship. SB is supported by the UK Research and Innovation (UKRI) Future Leaders Fellowship (MR/V023381/1). BL is supported by STFC via Consolidated Grants ST/I00162X/1, ST/P000541/1 and ST/X001075/1 for this project. We thank David Weinberg for his insightful comments. 

The simulations and analyses of this project made use of the DiRAC@Durham facility managed by the Institute for Computational Cosmology (ICC) on behalf of the STFC DiRAC HPC Facility (\href{www.dirac.ac.uk}{www.dirac.ac.uk}). The equipment was funded by BEIS capital funding via STFC capital grants ST/K00042X/1, ST/P002293/1, ST/R002371/1 and ST/S002502/1, Durham University and STFC operations grant ST/R000832/1. DiRAC is part of the National e-Infrastructure.

\section*{Data Availability}

The galaxy catalogues used in this paper may be requested from the authors.




\bibliographystyle{mnras}
\bibliography{RSD_ST_MG} 

\begin{thebibliography}{}
\makeatletter
\relax
\def\mn@urlcharsother{\let\do\@makeother \do\$\do\&\do\#\do\^\do\_\do\%\do\~}
\def\mn@doi{\begingroup\mn@urlcharsother \@ifnextchar [ {\mn@doi@} {\mn@doi@[]}}
\def\mn@doi@[#1]#2{\def\@tempa{#1}\ifx\@tempa\@empty \href {http://dx.doi.org/#2} {doi:#2}\else \href {http://dx.doi.org/#2} {#1}\fi \endgroup}
\def\mn@eprint#1#2{\mn@eprint@#1:#2::\@nil}
\def\mn@eprint@arXiv#1{\href {http://arxiv.org/abs/#1} {{\tt arXiv:#1}}}
\def\mn@eprint@dblp#1{\href {http://dblp.uni-trier.de/rec/bibtex/#1.xml} {dblp:#1}}
\def\mn@eprint@#1:#2:#3:#4\@nil{\def\@tempa {#1}\def\@tempb {#2}\def\@tempc {#3}\ifx \@tempc \@empty \let \@tempc \@tempb \let \@tempb \@tempa \fi \ifx \@tempb \@empty \def\@tempb {arXiv}\fi \@ifundefined {mn@eprint@\@tempb}{\@tempb:\@tempc}{\expandafter \expandafter \csname mn@eprint@\@tempb\endcsname \expandafter{\@tempc}}}

\bibitem[\protect\citeauthoryear{Abbott et~al.,}{Abbott et~al.}{2016}]{Abbott_et-al_2016}
Abbott B.~P.,  et~al., 2016, \mn@doi [Phys. Rev. Lett.] {10.1103/PhysRevLett.116.061102}, 116, 061102

\bibitem[\protect\citeauthoryear{Abbott et~al.,}{Abbott et~al.}{2022}]{DES_1}
Abbott T. M.~C.,  et~al., 2022, \mn@doi [Phys. Rev. D] {10.1103/PhysRevD.105.023520}, 105, 023520

\bibitem[\protect\citeauthoryear{Arnold, Springel  \& Puchwein}{Arnold et~al.}{2016}]{Arnold_2016}
Arnold C.,  Springel V.,   Puchwein E.,  2016, \mn@doi [Monthly Notices of the Royal Astronomical Society] {10.1093/mnras/stw1708}, 462, 1530

\bibitem[\protect\citeauthoryear{Arnold, Leo  \& Li}{Arnold et~al.}{2019}]{Arnold_2019}
Arnold C.,  Leo M.,   Li B.,  2019, \mn@doi [Nature Astronomy] {10.1038/s41550-019-0823-y}, 3, 945–954

\bibitem[\protect\citeauthoryear{{Babichev}, {Deffayet}  \& {Ziour}}{{Babichev} et~al.}{2009}]{Babichev:2009IJMPD..18.2147B}
{Babichev} E.,  {Deffayet} C.,   {Ziour} R.,  2009, \mn@doi [International Journal of Modern Physics D] {10.1142/S0218271809016107}, \href {https://ui.adsabs.harvard.edu/abs/2009IJMPD..18.2147B} {18, 2147}

\bibitem[\protect\citeauthoryear{{Baugh}}{{Baugh}}{2006}]{2006RPPh...69.3101B}
{Baugh} C.~M.,  2006, \mn@doi [Reports on Progress in Physics] {10.1088/0034-4885/69/12/R02}, \href {https://ui.adsabs.harvard.edu/abs/2006RPPh...69.3101B} {69, 3101}

\bibitem[\protect\citeauthoryear{Beltrán~Jiménez \& Heisenberg}{Beltrán~Jiménez \& Heisenberg}{2016}]{Beltr_n_Jim_nez_2016}
Beltrán~Jiménez J.,  Heisenberg L.,  2016, \mn@doi [Physics Letters B] {10.1016/j.physletb.2016.04.017}, 757, 405–411

\bibitem[\protect\citeauthoryear{Bennett et~al.,}{Bennett et~al.}{2013}]{Bennett_2013}
Bennett C.~L.,  et~al., 2013, \mn@doi [The Astrophysical Journal Supplement Series] {10.1088/0067-0049/208/2/20}, 208, 20

\bibitem[\protect\citeauthoryear{{Berlind} \& {Weinberg}}{{Berlind} \& {Weinberg}}{2002}]{2002ApJ...575..587B}
{Berlind} A.~A.,  {Weinberg} D.~H.,  2002, \mn@doi [\apj] {10.1086/341469}, \href {https://ui.adsabs.harvard.edu/abs/2002ApJ...575..587B} {575, 587}

\bibitem[\protect\citeauthoryear{Bose, Hellwing  \& Li}{Bose et~al.}{2015}]{Bose_2015}
Bose S.,  Hellwing W.~A.,   Li B.,  2015, \mn@doi [Journal of Cosmology and Astroparticle Physics] {10.1088/1475-7516/2015/02/034}, 2015, 034

\bibitem[\protect\citeauthoryear{{Brax}, {van de Bruck}, {Davis}  \& {Shaw}}{{Brax} et~al.}{2008}]{2008PhRvD..78j4021B}
{Brax} P.,  {van de Bruck} C.,  {Davis} A.-C.,   {Shaw} D.~J.,  2008, \mn@doi [\prd] {10.1103/PhysRevD.78.104021}, \href {https://ui.adsabs.harvard.edu/abs/2008PhRvD..78j4021B} {78, 104021}

\bibitem[\protect\citeauthoryear{{Carroll}}{{Carroll}}{2001}]{CarrollLambda:2001LRR.....4....1C}
{Carroll} S.~M.,  2001, \mn@doi [Living Reviews in Relativity] {10.12942/lrr-2001-1}, \href {https://ui.adsabs.harvard.edu/abs/2001LRR.....4....1C} {4, 1}

\bibitem[\protect\citeauthoryear{Collaboration et~al.,}{Collaboration et~al.}{2024}]{desicollaboration2024}
Collaboration D.,  et~al., 2024, DESI 2024 VI: Cosmological Constraints from the Measurements of Baryon Acoustic Oscillations (\mn@eprint {arXiv} {2404.03002})

\bibitem[\protect\citeauthoryear{{Conroy} \& {Wechsler}}{{Conroy} \& {Wechsler}}{2009}]{Conroy_2009}
{Conroy} C.,  {Wechsler} R.~H.,  2009, \mn@doi [\apj] {10.1088/0004-637X/696/1/620}, \href {https://ui.adsabs.harvard.edu/abs/2009ApJ...696..620C} {696, 620}

\bibitem[\protect\citeauthoryear{Cuesta-Lazaro, Li, Eggemeier, Zarrouk, Baugh, Nishimichi  \& Takada}{Cuesta-Lazaro et~al.}{2020}]{Cuesta-Lazaro:2020ihk}
Cuesta-Lazaro C.,  Li B.,  Eggemeier A.,  Zarrouk P.,  Baugh C.~M.,  Nishimichi T.,   Takada M.,  2020, \mn@doi [Mon. Not. Roy. Astron. Soc.] {10.1093/mnras/staa2249}, 498, 1175

\bibitem[\protect\citeauthoryear{{DESI Collaboration Et Al}}{{DESI Collaboration Et Al}}{2023}]{DESI_EDR_2023}
{DESI Collaboration Et Al} 2023, The Early Data Release of the Dark Energy Spectroscopic Instrument, \mn@doi{10.5281/ZENODO.7964161}, \url {https://zenodo.org/record/7964161}

\bibitem[\protect\citeauthoryear{{De Felice} \& {Tsujikawa}}{{De Felice} \& {Tsujikawa}}{2010}]{DeFelice:2010LRR....13....3D}
{De Felice} A.,  {Tsujikawa} S.,  2010, \mn@doi [Living Reviews in Relativity] {10.12942/lrr-2010-3}, \href {https://ui.adsabs.harvard.edu/abs/2010LRR....13....3D} {13, 3}

\bibitem[\protect\citeauthoryear{Deffayet, Esposito-Farèse  \& Vikman}{Deffayet et~al.}{2009}]{Deffayet_2009}
Deffayet C.,  Esposito-Farèse G.,   Vikman A.,  2009, \mn@doi [Physical Review D] {10.1103/physrevd.79.084003}, 79

\bibitem[\protect\citeauthoryear{{Dvali}, {Gabadadze}  \& {Porrati}}{{Dvali} et~al.}{2000}]{DvaliDGP:2000PhLB..485..208D}
{Dvali} G.,  {Gabadadze} G.,   {Porrati} M.,  2000, \mn@doi [Physics Letters B] {10.1016/S0370-2693(00)00669-9}, \href {https://ui.adsabs.harvard.edu/abs/2000PhLB..485..208D} {485, 208}

\bibitem[\protect\citeauthoryear{{Efstathiou}}{{Efstathiou}}{2024}]{2024arXiv240612106E}
{Efstathiou} G.,  2024, \mn@doi [arXiv e-prints] {10.48550/arXiv.2406.12106}, \href {https://ui.adsabs.harvard.edu/abs/2024arXiv240612106E} {p. arXiv:2406.12106}

\bibitem[\protect\citeauthoryear{{Euclid Collaboration: Gabarra} et~al.,}{{Euclid Collaboration: Gabarra} et~al.}{2023}]{Gabarra_EP31}
{Euclid Collaboration: Gabarra} L.,  et~al., 2023, \mn@doi [aap] {10.48550/arXiv.2302.09372}, \href {https://ui.adsabs.harvard.edu/abs/2023arXiv230209372E} {p. arXiv:2302.09372}

\bibitem[\protect\citeauthoryear{Hadzhiyska, Bose, Eisenstein, Hernquist  \& Spergel}{Hadzhiyska et~al.}{2020}]{Hadzhiyska_2020}
Hadzhiyska B.,  Bose S.,  Eisenstein D.,  Hernquist L.,   Spergel D.~N.,  2020, \mn@doi [Monthly Notices of the Royal Astronomical Society] {10.1093/mnras/staa623}, 493, 5506–5519

\bibitem[\protect\citeauthoryear{Hadzhiyska, Tacchella, Bose  \& Eisenstein}{Hadzhiyska et~al.}{2021}]{Hadzhiyska_2021}
Hadzhiyska B.,  Tacchella S.,  Bose S.,   Eisenstein D.~J.,  2021, \mn@doi [Monthly Notices of the Royal Astronomical Society] {10.1093/mnras/stab243}, 502, 3599–3617

\bibitem[\protect\citeauthoryear{Heisenberg}{Heisenberg}{2014}]{Heisenberg_2014}
Heisenberg L.,  2014, \mn@doi [Journal of Cosmology and Astroparticle Physics] {10.1088/1475-7516/2014/05/015}, 2014, 015–015

\bibitem[\protect\citeauthoryear{Hernández-Aguayo, Ruan, Li, Arnold, Baugh, Klypin  \& Prada}{Hernández-Aguayo et~al.}{2022}]{Hern_ndez_Aguayo_2022}
Hernández-Aguayo C.,  Ruan C.-Z.,  Li B.,  Arnold C.,  Baugh C.~M.,  Klypin A.,   Prada F.,  2022, \mn@doi [Journal of Cosmology and Astroparticle Physics] {10.1088/1475-7516/2022/01/048}, 2022, 048

\bibitem[\protect\citeauthoryear{Heymans et~al.,}{Heymans et~al.}{2021}]{KiDS_1}
Heymans C.,  et~al., 2021, \mn@doi [A&A] {10.1051/0004-6361/202039063}, 646, A140

\bibitem[\protect\citeauthoryear{Hikage et~al.,}{Hikage et~al.}{2019}]{HSC_1}
Hikage C.,  et~al., 2019, \mn@doi [Publications of the Astronomical Society of Japan] {10.1093/pasj/psz010}, 71, 43

\bibitem[\protect\citeauthoryear{{Hu} \& {Sawicki}}{{Hu} \& {Sawicki}}{2007}]{Hu:2007PhRvD..76f4004H}
{Hu} W.,  {Sawicki} I.,  2007, \mn@doi [\prd] {10.1103/PhysRevD.76.064004}, \href {https://ui.adsabs.harvard.edu/abs/2007PhRvD..76f4004H} {76, 064004}

\bibitem[\protect\citeauthoryear{{Jennings}, {Baugh}, {Li}, {Zhao}  \& {Koyama}}{{Jennings} et~al.}{2012}]{2012MNRAS.425.2128J}
{Jennings} E.,  {Baugh} C.~M.,  {Li} B.,  {Zhao} G.-B.,   {Koyama} K.,  2012, \mn@doi [\mnras] {10.1111/j.1365-2966.2012.21567.x}, \href {https://ui.adsabs.harvard.edu/abs/2012MNRAS.425.2128J} {425, 2128}

\bibitem[\protect\citeauthoryear{Khoury \& Weltman}{Khoury \& Weltman}{2004a}]{Khoury_2004b}
Khoury J.,  Weltman A.,  2004a, \mn@doi [Phys. Rev. D] {10.1103/PhysRevD.69.044026}, 69, 044026

\bibitem[\protect\citeauthoryear{Khoury \& Weltman}{Khoury \& Weltman}{2004b}]{Khoury_2004}
Khoury J.,  Weltman A.,  2004b, \mn@doi [Phys. Rev. Lett.] {10.1103/PhysRevLett.93.171104}, 93, 171104

\bibitem[\protect\citeauthoryear{{Kravtsov}, {Berlind}, {Wechsler}, {Klypin}, {Gottl{\"o}ber}, {Allgood}  \& {Primack}}{{Kravtsov} et~al.}{2004}]{Kravtsov_2004}
{Kravtsov} A.~V.,  {Berlind} A.~A.,  {Wechsler} R.~H.,  {Klypin} A.~A.,  {Gottl{\"o}ber} S.,  {Allgood} B.,   {Primack} J.~R.,  2004, \mn@doi [\apj] {10.1086/420959}, \href {https://ui.adsabs.harvard.edu/abs/2004ApJ...609...35K} {609, 35}

\bibitem[\protect\citeauthoryear{Li, Zhao, Teyssier  \& Koyama}{Li et~al.}{2012}]{Li_2012}
Li B.,  Zhao G.-B.,  Teyssier R.,   Koyama K.,  2012, \mn@doi [Journal of Cosmology and Astroparticle Physics] {10.1088/1475-7516/2012/01/051}, 2012, 051–051

\bibitem[\protect\citeauthoryear{{Li}, {Hellwing}, {Koyama}, {Zhao}, {Jennings}  \& {Baugh}}{{Li} et~al.}{2013}]{2013MNRAS.428..743L}
{Li} B.,  {Hellwing} W.~A.,  {Koyama} K.,  {Zhao} G.-B.,  {Jennings} E.,   {Baugh} C.~M.,  2013, \mn@doi [\mnras] {10.1093/mnras/sts072}, \href {https://ui.adsabs.harvard.edu/abs/2013MNRAS.428..743L} {428, 743}

\bibitem[\protect\citeauthoryear{{Mitchell} et~al.,}{{Mitchell} et~al.}{2018a}]{2018MNRAS.474..492M}
{Mitchell} P.~D.,  et~al., 2018a, \mn@doi [\mnras] {10.1093/mnras/stx2770}, \href {https://ui.adsabs.harvard.edu/abs/2018MNRAS.474..492M} {474, 492}

\bibitem[\protect\citeauthoryear{Mitchell, He, Arnold  \& Li}{Mitchell et~al.}{2018b}]{Mitchell_2018}
Mitchell M.~A.,  He J.-h.,  Arnold C.,   Li B.,  2018b, \mn@doi [Monthly Notices of the Royal Astronomical Society] {10.1093/mnras/sty636}, 477, 1133

\bibitem[\protect\citeauthoryear{Mitchell, Arnold  \& Li}{Mitchell et~al.}{2022}]{Mitchell_2022}
Mitchell M.~A.,  Arnold C.,   Li B.,  2022, \mn@doi [Monthly Notices of the Royal Astronomical Society] {10.1093/mnras/stac1528}, 514, 3349–3365

\bibitem[\protect\citeauthoryear{{Mota} \& {Shaw}}{{Mota} \& {Shaw}}{2006}]{2006PhRvL..97o1102M}
{Mota} D.~F.,  {Shaw} D.~J.,  2006, \mn@doi [\prl] {10.1103/PhysRevLett.97.151102}, \href {https://ui.adsabs.harvard.edu/abs/2006PhRvL..97o1102M} {97, 151102}

\bibitem[\protect\citeauthoryear{Nicolis, Rattazzi  \& Trincherini}{Nicolis et~al.}{2009}]{Nicolis_2009}
Nicolis A.,  Rattazzi R.,   Trincherini E.,  2009, \mn@doi [Physical Review D] {10.1103/physrevd.79.064036}, 79

\bibitem[\protect\citeauthoryear{{Perlmutter} et~al.,}{{Perlmutter} et~al.}{1999}]{1999ApJ...517..565P}
{Perlmutter} S.,  et~al., 1999, \mn@doi [\apj] {10.1086/307221}, \href {https://ui.adsabs.harvard.edu/abs/1999ApJ...517..565P} {517, 565}

\bibitem[\protect\citeauthoryear{Pillepich et~al.}{Pillepich et~al.}{2018}]{Pillepich:2017jle}
Pillepich A.,  et~al., 2018, \mn@doi [Mon. Not. Roy. Astron. Soc.] {10.1093/mnras/stx2656}, 473, 4077

\bibitem[\protect\citeauthoryear{{Planck Collaboration} et~al.,}{{Planck Collaboration} et~al.}{2016a}]{Planck15Overview:2016A&A...594A...1P}
{Planck Collaboration} et~al., 2016a, \mn@doi [\aap] {10.1051/0004-6361/201527101}, \href {https://ui.adsabs.harvard.edu/abs/2016A&A...594A...1P} {594, A1}

\bibitem[\protect\citeauthoryear{{Planck Collaboration} et~al.,}{{Planck Collaboration} et~al.}{2016b}]{Planck15Parameters:2016A&A...594A..13P}
{Planck Collaboration} et~al., 2016b, \mn@doi [\aap] {10.1051/0004-6361/201525830}, \href {https://ui.adsabs.harvard.edu/abs/2016A&A...594A..13P} {594, A13}

\bibitem[\protect\citeauthoryear{Puchwein, Baldi  \& Springel}{Puchwein et~al.}{2013}]{Puchwein_2013}
Puchwein E.,  Baldi M.,   Springel V.,  2013, \mn@doi [Monthly Notices of the Royal Astronomical Society] {10.1093/mnras/stt1575}, 436, 348

\bibitem[\protect\citeauthoryear{Reverberi \& Daverio}{Reverberi \& Daverio}{2019}]{Reverberi_2019}
Reverberi L.,  Daverio D.,  2019, \mn@doi [Journal of Cosmology and Astroparticle Physics] {10.1088/1475-7516/2019/07/035}, 2019, 035–035

\bibitem[\protect\citeauthoryear{{Riess} et~al.,}{{Riess} et~al.}{1998}]{1998AJ....116.1009R}
{Riess} A.~G.,  et~al., 1998, \mn@doi [\aj] {10.1086/300499}, \href {https://ui.adsabs.harvard.edu/abs/1998AJ....116.1009R} {116, 1009}

\bibitem[\protect\citeauthoryear{{Ruan}, {Hern{\'a}ndez-Aguayo}, {Li}, {Arnold}, {Baugh}, {Klypin}  \& {Prada}}{{Ruan} et~al.}{2021}]{Ruan:2021MGGLAMfR}
{Ruan} C.-Z.,  {Hern{\'a}ndez-Aguayo} C.,  {Li} B.,  {Arnold} C.,  {Baugh} C.~M.,  {Klypin} A.,   {Prada} F.,  2021, arXiv e-prints, \href {https://ui.adsabs.harvard.edu/abs/2021arXiv211000328R} {p. arXiv:2110.00328}

\bibitem[\protect\citeauthoryear{{Ruan}, {Cuesta-Lazaro}, {Eggemeier}, {Hern{\'a}ndez-Aguayo}, {Baugh}, {Li}  \& {Prada}}{{Ruan} et~al.}{2022a}]{2022MNRAS.514..440R}
{Ruan} C.-Z.,  {Cuesta-Lazaro} C.,  {Eggemeier} A.,  {Hern{\'a}ndez-Aguayo} C.,  {Baugh} C.~M.,  {Li} B.,   {Prada} F.,  2022a, \mn@doi [\mnras] {10.1093/mnras/stac1345}, \href {https://ui.adsabs.harvard.edu/abs/2022MNRAS.514..440R} {514, 440}

\bibitem[\protect\citeauthoryear{Ruan, Hernández-Aguayo, Li, Arnold, Baugh, Klypin  \& Prada}{Ruan et~al.}{2022b}]{Ruan_2022}
Ruan C.-Z.,  Hernández-Aguayo C.,  Li B.,  Arnold C.,  Baugh C.~M.,  Klypin A.,   Prada F.,  2022b, \mn@doi [Journal of Cosmology and Astroparticle Physics] {10.1088/1475-7516/2022/05/018}, 2022, 018

\bibitem[\protect\citeauthoryear{Sotiriou \& Faraoni}{Sotiriou \& Faraoni}{2010}]{Sotiriou_2010}
Sotiriou T.~P.,  Faraoni V.,  2010, \mn@doi [Reviews of Modern Physics] {10.1103/revmodphys.82.451}, 82, 451–497

\bibitem[\protect\citeauthoryear{{Springel}}{{Springel}}{2010}]{2010MNRAS.401..791S}
{Springel} V.,  2010, \mn@doi [\mnras] {10.1111/j.1365-2966.2009.15715.x}, \href {https://ui.adsabs.harvard.edu/abs/2010MNRAS.401..791S} {401, 791}

\bibitem[\protect\citeauthoryear{Sullivan et~al.,}{Sullivan et~al.}{2011}]{Sullivan_2011}
Sullivan M.,  et~al., 2011, \mn@doi [The Astrophysical Journal] {10.1088/0004-637x/737/2/102}, 737, 102

\bibitem[\protect\citeauthoryear{Tasitsiomi, Kravtsov, Wechsler  \& Primack}{Tasitsiomi et~al.}{2004}]{Tasitsiomi_2004}
Tasitsiomi A.,  Kravtsov A.~V.,  Wechsler R.~H.,   Primack J.~R.,  2004, \mn@doi [The Astrophysical Journal] {10.1086/423784}, 614, 533

\bibitem[\protect\citeauthoryear{Vale \& Ostriker}{Vale \& Ostriker}{2004}]{Vale_2004}
Vale A.,  Ostriker J.~P.,  2004, \mn@doi [Monthly Notices of the Royal Astronomical Society] {10.1111/j.1365-2966.2004.08059.x}, 353, 189–200

\bibitem[\protect\citeauthoryear{Wechsler \& Tinker}{Wechsler \& Tinker}{2018}]{Wechsler_2018}
Wechsler R.~H.,  Tinker J.~L.,  2018, \mn@doi [Annual Review of Astronomy and Astrophysics] {10.1146/annurev-astro-081817-051756}, 56, 435–487

\bibitem[\protect\citeauthoryear{{Yuan}, {Hadzhiyska}, {Bose}  \& {Eisenstein}}{{Yuan} et~al.}{2022}]{2022MNRAS.512.5793Y}
{Yuan} S.,  {Hadzhiyska} B.,  {Bose} S.,   {Eisenstein} D.~J.,  2022, \mn@doi [\mnras] {10.1093/mnras/stac830}, \href {https://ui.adsabs.harvard.edu/abs/2022MNRAS.512.5793Y} {512, 5793}

\makeatother
\end{thebibliography}



\appendix

\section{The Halo Occupation Distributions}
\label{appendix:HOD}

In this appendix we show the general behaviour of the HODs of the galaxy catalogues whose clustering was studied in detail in the main text. This will help us better understand the clustering signals seen above.

We will analyse the HOD in greater detail, including providing fitting formulae for both LRGs and ELGs, at different redshifts and number densities, in a future work.

\subsection{The behaviour of ELG HOD}

\begin{figure*}
    \centering
    \includegraphics[scale = 0.69]{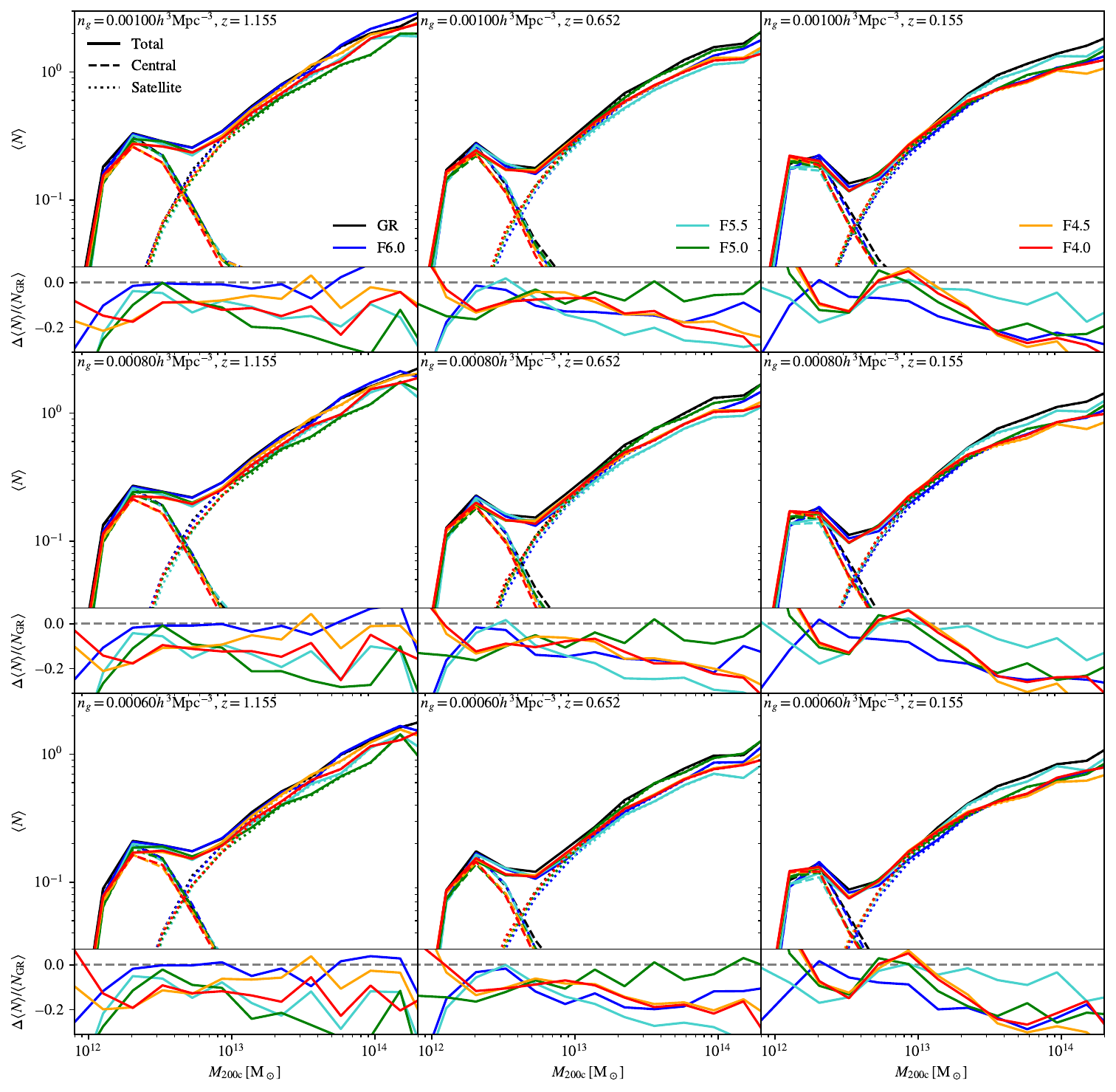}
    \caption{\textit{Large panels}: The mean ELG occupations of halos as a function of the dark matter mass within $R_{200c}$, the mean total occupations, satellite galaxy occupations and central galaxy occupations are all shown with separate lines. \textit{Small panels:} The corresponding fractional difference between model and GR mean total ELG occupations. Each large upper and small lower panel pair corresponds to ELG samples produced at a different number density-redshift pair stated at the top of the large panel.}
    \label{fig:ELG HOD}
\end{figure*}

The HODs for the ELG catalogues are shown in Fig.~\ref{fig:ELG HOD}, where we note that relative differences between models, like the CFs in Fig.~\ref{fig:ELG real space cf}, depend very little on the number density, with the lower fractional difference panels at the same redshift all looking visually very similar and the models having the same relative behaviour; as such, we can focus on just the redshift dependency. 

This fits well with the observation above that weaker MG models are ``delayed'' versions of stronger ones, and the narrative goes as follows: 
\begin{enumerate}
    \item at some earlier time, small haloes in a MG model become unscreened and the abundance of haloes at that mass receives a boost, and the mean halo occupation there dips as the total galaxy number in our catalogues is fixed. At this point, more massive haloes are still screened, so that their HODs remain the same as in GR.
    \item next, the unscreened small haloes have had long enough time to accrete and cool more gas to form young star-forming galaxies. This means that some of the galaxies in those small haloes can have higher specific SFR than some galaxies in larger haloes; because our ELG catalogue has been selected by sorting the sSFR from high to low, this means that more of the ELGs will be from smaller haloes, and for a fixed number density this can only imply that fewer will be from larger haloes. This leads to a reduction to the halo occupacy number in the latter.
    \item as times goes on, larger haloes progressively get unscreened, and the halo occupancy number increases again in these haloes. 
    \item finally, quenching of star formation comes to effect, which further suppresses the halo occupancy number.
\end{enumerate}

As an example, looking at the top panels of Fig.~\ref{fig:ELG HOD}: at $z=1.155$ F6.0 is at stage (i), F5.5 and F5.0 at stage (ii), F4.5 at stage (iii) and F4.0 at stage (iv); at $z=0.652$ F6.0 and F5.5 are both at stage (ii), F5.0 at stage (iii), and F4.5 and F4.0 at stage (iv); at $z=0.155$ F6.0 is at stage (ii), F5.5 at stage (iii) and F5.0, F4.5, F4.0 at stage (iv). 

It should be evident that stage (ii) is the most interesting one, as this is when more ELGs are hosted by smaller haloes (or more accurately speaking, haloes forming from lower initial density peaks) which are inherently less clustered. This explains well why the MG model with the weakest clustering is F5.0 at $z=1.155$, F5.5 at $z=0.652$ and F6.0 at $z=0.155$, cf.~Fig.~\ref{fig:ELG real space cf}. 

Therefore, this reiterates the important point that the specific galaxy type is important for determining the strength of the clustering signal. With suitable choices of galaxies, such as ELGs which tend to be hosted by smaller haloes that get unscreened early and more easily, one can hope to maximise the potential of distinguishing between different gravity models. Even the weakest MG models, such as F6.0, can be hopefully distinguishable from GR by looking at the ``correct" redshifts.

Finally, we notice that at low redshift the HOD plots also display the ``saturation" effect mentioned above. This can be seen by observing that, at $z=0.652$ and $0.155$, F4.0 and F4.5 have nearly identical HODs; indeed, at $z=0.155$ even F5.0 has nearly the same HOD as F4.5 and F4.0. 

\subsection{The behaviour of LRG HOD}

\begin{figure*}
    \centering
    \includegraphics[scale = 0.69]{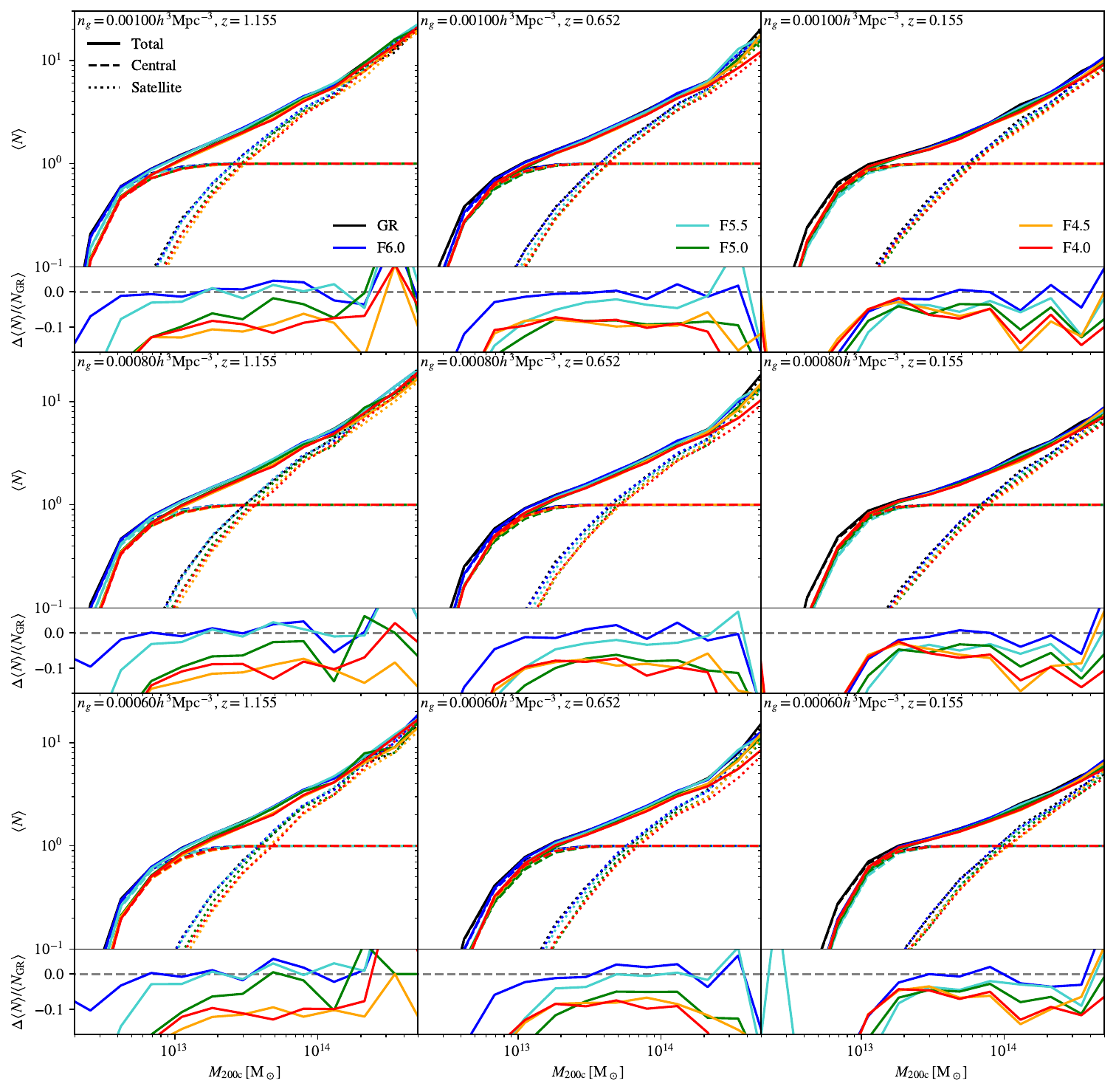}
    \caption{\textit{Large panels}: The mean LRG occupations of halos as a function of the dark matter mass within $R_{200c}$, the mean total occupations, satellite galaxy occupations and central galaxy occupations are all shown with separate lines. \textit{Small panels:} The corresponding fractional difference between model and GR mean total LRG occupations. Each large upper and small lower panel pair corresponds to LRG samples produced at a different number density-redshift pair stated at the top of the large panel.}
    \label{fig:LRG HOD}
\end{figure*}

The HOD for LRGs is shown in Fig.~\ref{fig:LRG HOD}, which displays much simpler behaviour. Above some mass, virtually all haloes host a central galaxy (the halo occupancy number is $1.0$); below this mass there is a transitional regime in which the the occupations fall towards zero. The cutoff mass where the halo occupation starts to drop is always smaller for GR. The reason for this can be found by looking at the HMFs. We have an excess of haloes above the equivalent GR cutoff mass for the HOD in all models, so for the same number of galaxies the cutoff mass is higher in MG models.

In terms of the deviations from GR, unlike ELGs, the HOD for LRGs shows a well defined trend in order of the strength of the fifth force. For example, F6.0 shows nearly identical HOD as GR except in the smallest host haloes; F5.5 shows a slight deviation from GR, while the deviations from GR by F5.0, F4.5 and F4.0 gets progressively stronger. Indeed the same ``saturation" effect can be observed at $z=0.652$ and $0.155$ where F4.5 and F4.0 are again nearly identical.

The HODs in MG models are generally lower than those in GR, and at the high-mass end this is mainly driven by satellite galaxies: the same density peaks in the initial condition lead to more massive haloes in MG models than in GR, but the numbers of satellite LRGs in these haloes are not necessarily or sufficiently enhanced in MG, and the consequence of this is a shift of the satellite occupation number curve to the right in MG models.

Overall, the HODs in our LRG catalogues confirm that the selection effect of host haloes is less prominent than for ELGs. While the LRG-hosting halo populations are still different between the different gravity models, this difference is mainly in the high-mass end which is not particularly beneficial if one wishes to constrain the weaker MG models.
    
\bsp	
\label{lastpage}
\end{document}